\documentclass[aps,rmp,reprint,amsmath,amssymb,graphicx,longbibliography,superscriptaddress]{revtex4-1}
\usepackage{graphicx}
\usepackage{xcolor}
\usepackage{ulem}
\usepackage{bm}
\usepackage{slashed}

\newcommand{\be}{\begin{equation}}
\newcommand{\ee}{\end{equation}}
\newcommand{\ba}{\begin{eqnarray}}
\newcommand{\ea}{\end{eqnarray}}

\newcommand{\red}[1]{{\color{red} #1}}

\newcommand{\uncomment}[1]{{\red{[ .... ]}}}

\begin{document}
\setcounter{page}{0} 
\title{Nucleon resonances and transition form factors}
\author {\large V.~D.~Burkert}
\affiliation{\large Thomas Jefferson National Accelerator Facility,\unpenalty~Newport News,\unpenalty~VA,\unpenalty~USA}
\date{\today}
\maketitle

\subsection{Abstract}
\noindent This is a contribution to the review "50 Years of Quantum Chromodynamics" edited by F. Gross and E. Klempt, to be published in Journal EPJC. This contribution reviews the nucleon resonance transition form factors determined from meson electro-production experiments at electron accelerator facilities, i.e. this contribution focuses on "space-like" transition form factors and amplitudes. Comparisons are made when available to LQCD and to approaches with traceable links to strong QCD and to advanced  quark model calculations.

\subsection{Introduction and Formalism}
\noindent Meson photoproduction has become an essential tool in the search for new excited light-quark baryon states. As discussed in the previous section, many new excited states have been discovered thanks to high precision photoproduction data in different final states~\cite{Anisovich:2017bsk}, and are now included in recent editions of the Review of Particle Physics (RPP)~\cite{Workman:2022ynf}. The exploration of the internal structure 
of excited states and the effective degrees of freedom contributing to s-channel resonance excitation requires the use of electron beams, which is the subject of this contribution, 
where the virtuality ($Q^2$) of the exchanged photon can be varied to pierce through the peripheral meson cloud and probe the quark core and its spatial structure. Electroproduction can thus say something about if a resonance is generated through short distance photon interaction with the small quark core, or through interaction with a more extended hadronic system. 

The experimental exploration of resonance transition form factors reaches over 60 years with many 
review articles describing this history. Here we refer to a few recent ones~\cite{Stoler:1993yk,Burkert:2004sk,Aznauryan:2011qj,Aznauryan:2012ba}. A review of recent electroproduction experiments in hadron physics and their interpretation within modern approaches of strong interaction physics can be found in Ref.~\cite{Proceedings:2020fyd}.       

\begin{figure}[h!]
\resizebox{1.0\columnwidth}{!}{\includegraphics{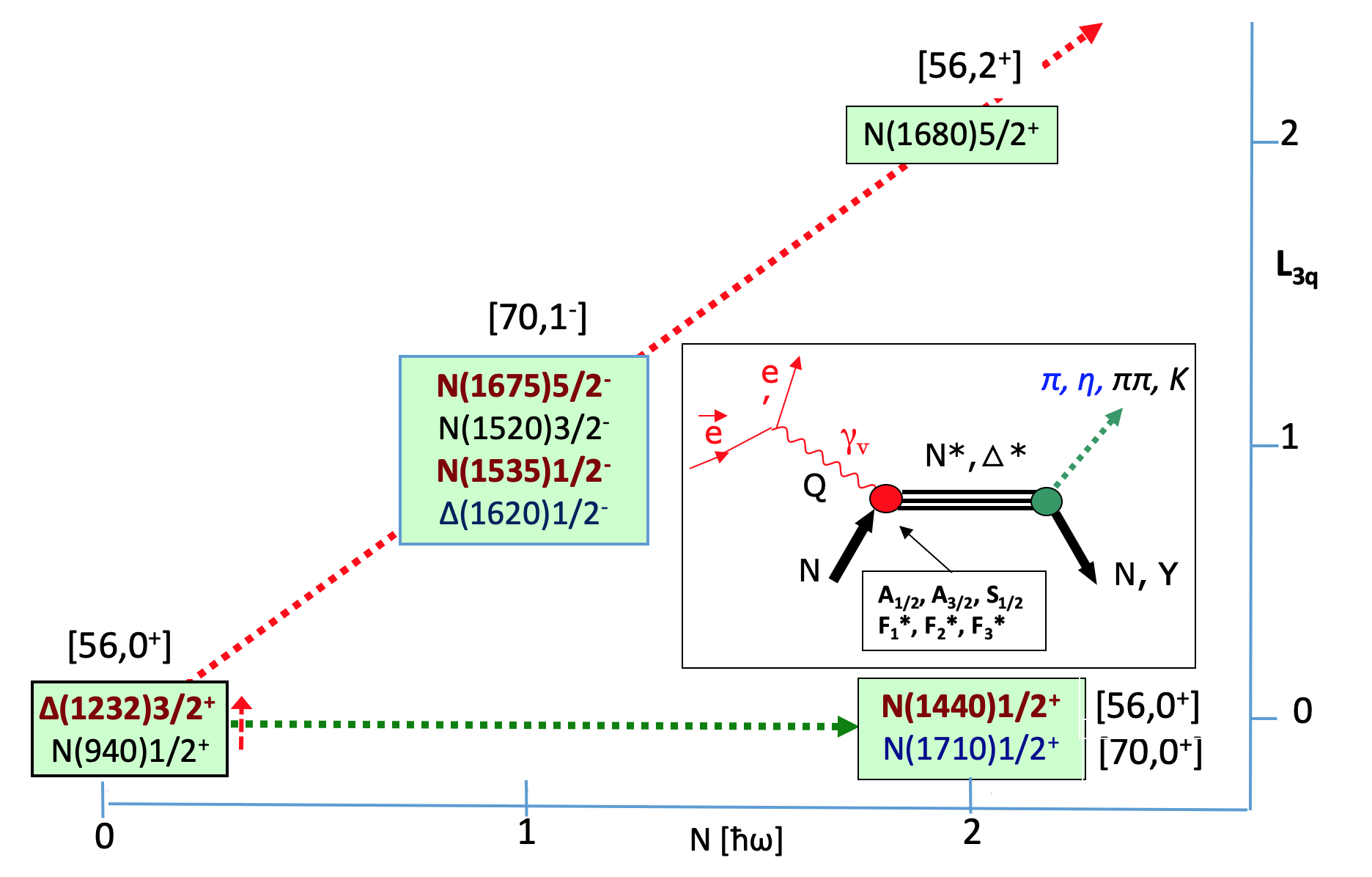}}
\caption{Excited states of the proton that have been studied in electroproduction to determine their resonance transition amplitudes or form factors. States highlighted in red are discussed in this subsection. Graphics from Ref.~\cite{Burkert:2019kxy}.}
\label{SU6}
\end{figure}

Electroproduction of final states with pseudoscalar mesons 
(e.g. $N\pi$, $p\eta$, $K\Lambda$) have been employed at Jefferson Laboratory mostly with the CEBAF Large Acceptance Spectrometer (CLAS) operating at an instantaneous luminosity of $10^{34}$sec$^{-1}$cm$^{-2}$. In Hall A and Hall C, pairs of individual well-shielded focusing magnetic spectrometers are employed with more specialized aims and limited acceptance, but operating at much higher luminosity. This experimental program led to new insights into the scale dependence of effective degrees of freedom, e.g. meson-baryon, constituent quarks, and dressed quark contributions. Several excited states, shown in Fig.~\ref{SU6} assigned to their primary $SU(6) \otimes O(3)$ supermultiplets, have been studied this way, mostly with CLAS in Hall B.  
Most of the resonance couplings have been extracted from single pseudoscalar meson production. In electroproduction, there are 6 complex helicity amplitudes, requiring a minimum of 11 
independent measurements for a complete~\footnote{With the exception of an overall phase that cannot be determined} model-independent determination of the amplitudes. In addition, measurements of isospin amplitudes
require additional measurements. Following this, the complex amplitudes would need to be subjected to analyses of their 
phase motions to determine resonance masses on the (real) energy axis, or poles in the (complex) energy plane.        
Fortunately, in the lower mass range a variety of constraints  can be applied to limit the number of unknowns when fitting the cross section data. 
These include the masses of states quite well known from hadronic processes or from meson photoproduction. Also, the number of possible
angular momenta is limited to $l_\pi \le 2$ in the examples discussed in the following. Additional constraints come from the Watson theorem~\cite{Watson:1954uc} that 
relates the electromagnetic phases to the hadronic ones, and the use of dispersion relations, assuming the imaginary parts of the amplitude are 
given by the resonance contribution, and the real parts determined through dispersion integrals and additional pole terms. Other approaches 
use unitary isobar models that parameterize all known resonances and background terms, and unitarize the full amplitudes in a K-matrix
procedure.  In the following, we show results based on both approaches, where additional systematic uncertainties have been derived from the 
differences in the two procedures.                

The availability of electron accelerators with the possibility of generating high beam currents at CEBAF at Jefferson Lab in the US and MAMI at Mainz in Germany, has enabled precise studies of the internal structure of excited states in the $N^*$ and the $\Delta^*$ sectors employing s-channel 
resonance excitations in large ranges of photon virtuality $Q^2$. This has enabled probing the degrees of freedom relevant in the resonance 
excitation as a function of the distance scale probed. This is the topic we will elucidate in the following sections and the relevance to (approximations to) 
QCD. First we briefly discuss the formalism needed for a quantitative analysis of the single pseudoscalar meson electroproduction.  
\begin{figure}[ptb]
\hspace{-0.5cm}\resizebox{1.0\columnwidth}{!}{\includegraphics{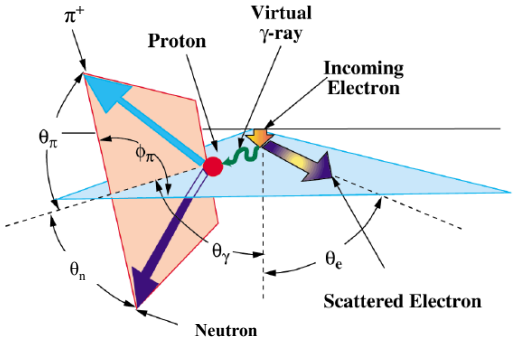}}
\caption{The kinematics of single $\pi^+$ electro-production off protons in the laboratory system.}
\label{kine}
\end{figure}
\subsubsection{Formalism in the analysis of electroproduction of single pseudoscalar mesons}
The simplest process used in the study of resonance transition amplitudes  is single pion or 
kaon production, e.g. $ep \to e\pi^+n$. Single $\pi^+$ and $\pi^0$ production are most 
suitable for the study of the lower-mass excited states as they couple dominantly to the 
excited states with masses up to $\approx 1.7$~GeV. It may then be useful to describe in more detail the 
analysis techniques and the formalism used.  
The unpolarized differential cross section for single pseudoscalar meson production 
can be written in the one-photon exchange approximation as: 
\begin{eqnarray}
 \frac {d\sigma} {dE_f d\Omega_e d\Omega_{\pi}} &=& \Gamma {\frac {d\sigma}  {d\Omega_{\pi}}}\,,  
\end{eqnarray}    
where $\Gamma$ is the virtual photon flux, 
\begin{eqnarray}
\Gamma = \frac{\alpha_{em}}{2\pi^2Q^2} \frac{(W^2-M^2)E_f}{2ME_i} \frac{1}{1 - \epsilon }, 
\end{eqnarray}
where $M$ is the proton mass, $W$ the mass of the hadronic final state, $\epsilon$ 
is the photon polarization parameter, $Q^2$ the photon virtuality, $E_i$ and $E_f$ represent the initial and the final electron energies, respectively. Moreover,
\begin{eqnarray} 
\epsilon = \left[ 1 + 2\left(1 + \frac{\nu^2}{Q^2}\right) \tan^2 \frac{\theta_e}{2}\right]^{-1}
\end{eqnarray} 
and
\begin{eqnarray}
\frac{d\sigma}{d\Omega_{\pi}} &=&\sigma_T + \epsilon \sigma_L + \epsilon\sigma_{TT}\cos2\phi_{\pi} \nonumber\\
&&\qquad\qquad\qquad + \sqrt{2\epsilon(1+\epsilon)}\sigma_{LT} \cos{\phi}_{\pi}\,. \nonumber
\end{eqnarray}
The kinematics for single $\pi^+$ production is shown in Fig.~\ref{kine}. 

\noindent The observables of the process $\gamma_v p \to \pi N'$ can be expressed in terms of six parity-conserving helicity amplitudes
~\cite{Aznauryan:2011qj,Walker:1968xu,Berends:1967vi} : 
 \begin{eqnarray}
{H_i = \left<\lambda_\pi;\lambda_N|T|\lambda_{\gamma_{\nu}} ; \lambda_p\right> , }
\end{eqnarray} 
where $\lambda$ denotes the helicity of the respective particle, $\lambda_\pi = 0$, $\lambda_N = \pm \frac{1}{2}$,  $\lambda_{\gamma_v} = \pm 1, 0$, and $\lambda_p = \pm \frac{1}{2}$, and $H_i$ are complex functions of $Q^2$, $W$, and $\theta^*_\pi$.  
\subsubsection{Multipoles and partial wave decompositions}
The response functions in (1) are given by: 
\begin{eqnarray}
\sigma_T &=& \frac{\vec{\it p}_\pi W}{2KM} (|H_1|^2 + |H_2|^2 + |H_3|^2 + |H_4|^2),\\
\sigma_L &=& \frac{\vec{\it p}_\pi W}{2KM} (|H_5|^2 + |H_6|^2),\\
\sigma_{TT} &=& \frac{\vec{\it p}_\pi W}{2KM} Re (H_2H_3^* - H_1H_4^*), \\
\sigma_{LT} &=& \frac{\vec{\it p}_\pi W}{2KM} Re [H_5^*(H_1-H_4) + H_6^* (H_2 + H_3)]\,,
\end{eqnarray} 
where $\vec{\it p}_\pi$ is the pion 3-momentum in the hadronic center-of-mass system, and $K$ is the equivalent real photon lab energy
needed to generate a state with mass $W$: 
\begin{eqnarray} 
K = \frac{W^2 - M^2}{2M}\,.
\end{eqnarray}  
The helicity amplitudes $H_i, i=1$--6, can be expanded into Legendre polynomials: 
\begin{eqnarray} 
H_1&=&\frac{1}{\sqrt{2}}\sin\theta\cos{\frac{\theta}{2}} \sum_{l=1}^\infty (B_{l+}-B_{(l+1)-})(P^{\prime\prime}_l-P^{\prime\prime}_{l+1})\nonumber
\end{eqnarray}
\begin{eqnarray}
H_2&=&\sqrt{2}\cos{\frac{\theta}{2}} \sum_{l=1}^\infty (A_{l+}-A_{(l+1)-})(P^{\prime}_l-P^{\prime}_{l+1})\nonumber\\
H_3&=&\frac{1}{\sqrt{2}}\sin\theta\sin{\frac{\theta}{2}} \sum_{l=1}^\infty (B_{l+}+B_{(l+1)-})(P^{\prime\prime}_l+P^{\prime\prime}_{l+1})\nonumber\\
H_4&=&\sqrt{2}\sin{\frac{\theta}{2}} \sum_{l=1}^\infty (A_{l+}+A_{(l+1)-})(P^{\prime}_l+P^{\prime}_{l+1})\nonumber\\
H_5&=&\sqrt{2}\cos{\frac{\theta}{2}} \sum_{l=1}^\infty (C_{l+}-C_{(l+1)-})(P^{\prime}_l-P^{\prime}_{l+1})\nonumber\\
H_6&=&\sqrt{2}\sin{\frac{\theta}{2}} \sum_{l=1}^\infty (C_{l+}+C_{(l+1)-})(P^{\prime}_l+P^{\prime}_{l+1})\,,
\end{eqnarray} 
where the $A_{l+}$ and $B_{l+}$ etc., are the transverse partial wave helicity elements for $\lambda_{\gamma p} =  \frac{1}{2}$  and 
$\lambda_{\gamma p} = \frac{3}{2}$, and $C_{\pm}$  the longitudinal partial wave helicity elements. 
In the subscript, $l+$ and $(l+1)-$ define the $\pi$ orbital angular momenta, and the sign
$\pm$ is related to the total angular momentum $J = l_{\pi} \pm \frac{1}{2}$.
In the analysis of data on the $N\Delta(1232)$ transition, linear combinations of partial wave helicity 
elements are expressed in terms of electromagnetic multipoles: 
\begin{eqnarray}
M_{l+} &=& \frac{1}{2(l+1)} [2A_{l+} - (l+2)B_{l+}] \\
E_{l+} &=&\frac{1}{2(l+1)} (2A_{l+} + lB_{l+}) \\
M_{l+1,-} &=& \frac{1}{2(l+1)} (2A_{l+1,-} - lB_{l+1,-}) \\
E_{l+1,-} &=&\frac{1}{2(l+1)} [-2A_{l+1,-} + (l+2)B_{l+1,-}] 
\end{eqnarray}
\begin{eqnarray}
S_{l+} &=& \frac{1}{ l+1} \sqrt{\frac{{\vec {\it Q^*}^2}}{Q^2}}C_{l+}  \\
S_{l+1,-} &=& \frac{1}{ l+1} \sqrt{\frac{{\vec {\it Q^*}^2}}{Q^2}}C_{l+1,-}\,,
\end{eqnarray} 
where $\vec{\it Q}^*$ is the photon 3-momentum in the hadronic rest frame. 
The electromagnetic multipoles are often used to describe the transition from the nucleon ground state to the $\Delta(1232)$, which is dominantly described as a magnetic dipole transition $M_{1+}$.      
The electromagnetic multipoles as well as the partial wave helicity elements are complex 
quantities and contain both non-resonant and resonant contributions. 
In order to compare the results to model predictions and LQCD, an additional analysis must 
be performed to separate the resonant parts $\hat{\rm A}_{\pm}$,  $\hat{\rm B}_{\pm}$, etc., 
from the non-resonant parts of the amplitudes. In a final step, the 
known hadronic properties of a given resonance can be used to determine 
photocoupling helicity amplitudes that characterize the  electromagnetic vertex:
\begin{eqnarray}
\hat{A}_{l\pm} &=& \mp FC^I_{\pi N} A_{1/2}, \\
\hat{B}_{l\pm} &=& \pm F \sqrt{\frac {16} {(2j -1)(2j + 3)} } C^I_{\pi N} A_{3/2}, \\
\hat{S}_{l\pm} &=& - F \frac{2\sqrt{2}}{2J+1} C^I_{\pi N} S_{1/2},  \\
F &=& \sqrt{\frac{1}{(2j + 1)}\pi} \frac {K}{p_\pi} \frac{\Gamma_\pi}{\Gamma^2} \nonumber
\end{eqnarray}
where the $C^I_{\pi N}$  are isospin coefficients.   
The total transverse absorption cross section for the transition into a specific resonance is given by: 
\begin{eqnarray}
\sigma_T = \frac{2M}{W_R\Gamma} (A^2_{1/2} + A^2_{3/2}). 
\end{eqnarray}

\noindent Experiments in the region of the $\Delta(1232)\frac{3}{2}^+$ resonance often determine the electric quadrupole ratio $R_{EM}$
\begin{equation}
R_{EM} = \frac{Im(E_{1+})}{Im(M_{1+})} \label{EM}
\end{equation} 
and the scalar quadrupole ratio $R_{SM}$ 
\begin{equation}
R_{SM} = \frac{Im(S_{1+})}{Im(M_{1+})} \label{SM}  
\end{equation} 
where $E_{1+}$, $S_{1+}$, and $M_{1+}$ are the electromagnetic transition multipoles at the mass of the $\Delta(1232)\frac{3}{2}^+$ resonance.   

It is worth noting that the electric and the scalar transition amplitudes are connected at the so-called pseudo-threshold $Q^2_{pt}=-(W-M)^2$ through the Siegert theorem~\cite{Siegert:1937yt}, whose impact on the extraction of nucleon resonance transition amplitudes from pion electro-production is discussed in~\cite{Tiator:2016kbr,Ramalho:2016zzo}.

\subsubsection{Resonance analysis tools} 

A model-independent determination of the amplitudes contributing to the electro-excitation of resonances in single pseudoscalar pion production $ep \to e^\prime N \pi$ (see kinematics  
of single pion production in Fig.~\ref{kine}) requires 
a large number of independent measurements at each value of the electron kinematics $W$, $Q^2$, the hadronic cms angle $\cos{\theta^\pi}$, and the azimuthal angle $\phi^\pi$ describing 
the angle between the electron scattering plane and the hadronic decay plane. Such a measurement
requires full exclusivity of the final state and employing both polarized electron beams and the measurements of the nucleon recoil polarization. 
\begin{figure}[th!]
\resizebox{1.0\columnwidth}{!}{\includegraphics{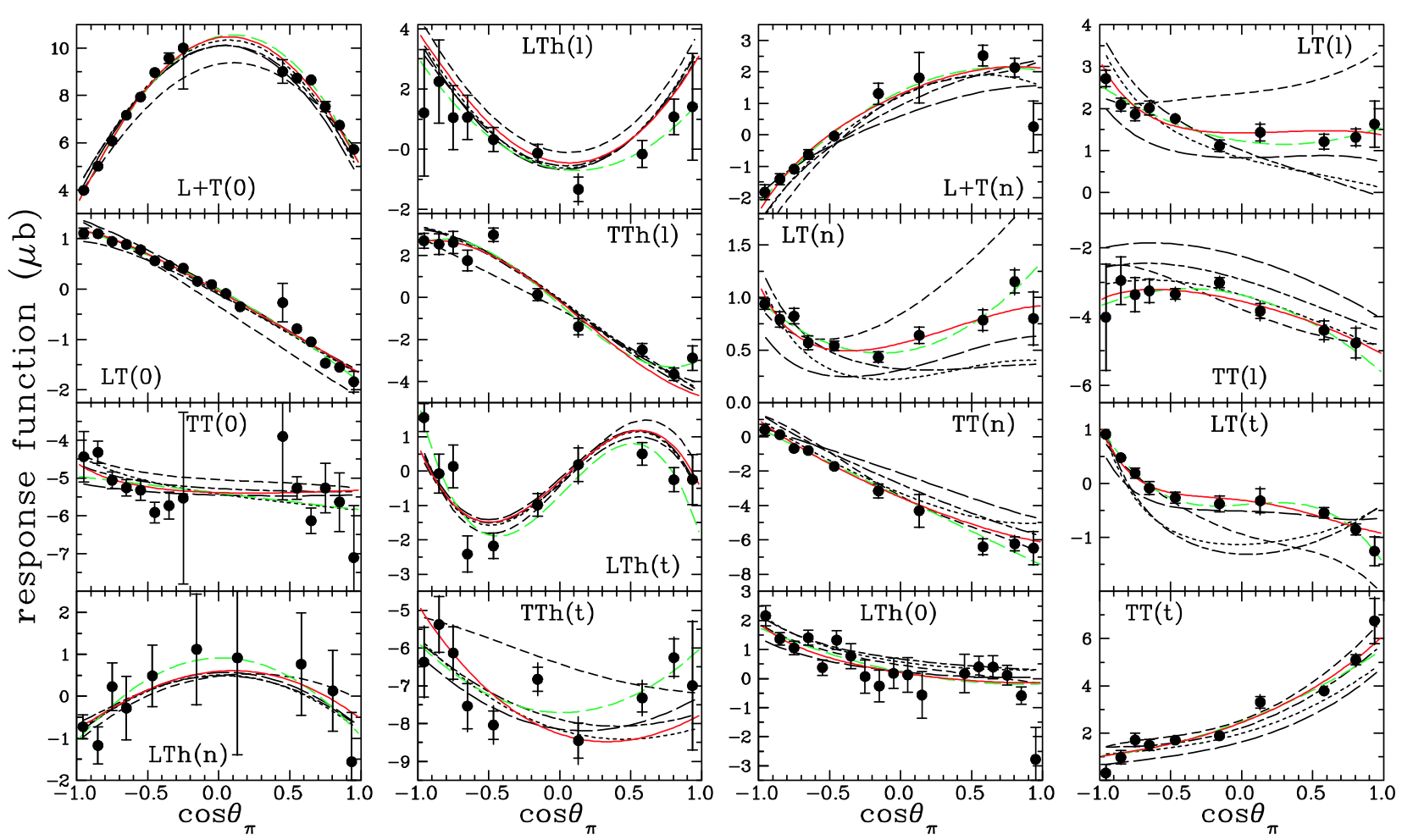}}
\caption{JLab/Hall A data for $\vec{e}p  \to e\vec{p}\pi^0$ response functions at W=1.232 GeV and $Q^2=1.0$~\cite{Kelly:2005jy}. Notations refer to transverse (t), normal (n) and longitudinal (l) components of the proton recoil polarization. The curves correspond to results obtained using SAID (short dashed), MAID (dashed-dotted), and the dynamical models DMT~\cite{Kamalov:2000en} (dotted), and SL~\cite{Sato:2000jf}   (long-dashed/green). The other curves correspond to Legendre and multipole fits performed by the authors.}
\label{Hall-A}
\end{figure}

Such measurements would in general require full $4\pi$ coverage for the hadronic final state. The only measurement that could claim to be complete was carried out at Jefferson Lab in Hall A~\cite{Kelly:2005jy} employing a limited kinematics centered at resonance  for $\vec{e} p \to e^{\prime} \vec{p} \pi^0$ at $W = 1.232$~GeV, and $Q^2 \approx 1$~GeV$^2$. Figure~\ref{Hall-A} shows the 16 response functions extracted from this measurement. The results of this measurement in terms of the magnetic $N\Delta$ transition form factor and the quadrupole ratios are included in 
Fig.~\ref{fig:Delta} among other data. They coincide very well with results of other experiments~\cite{CLAS:2009ces,CLAS:2006ezq,CLAS:2001cbm,Frolov:1998pw} using different analysis techniques that may be also applied to broader kinematic conditions, especially higher mass resonances. Details of the latter are discussed in ~\cite{Aznauryan:2011qj,Tiator:2011pw}. We briefly summarize them here:
\begin{figure*}[th!]
\centering
\begin{tabular}{ccc}
\includegraphics[width=0.30\textwidth,height=0.30\textwidth]{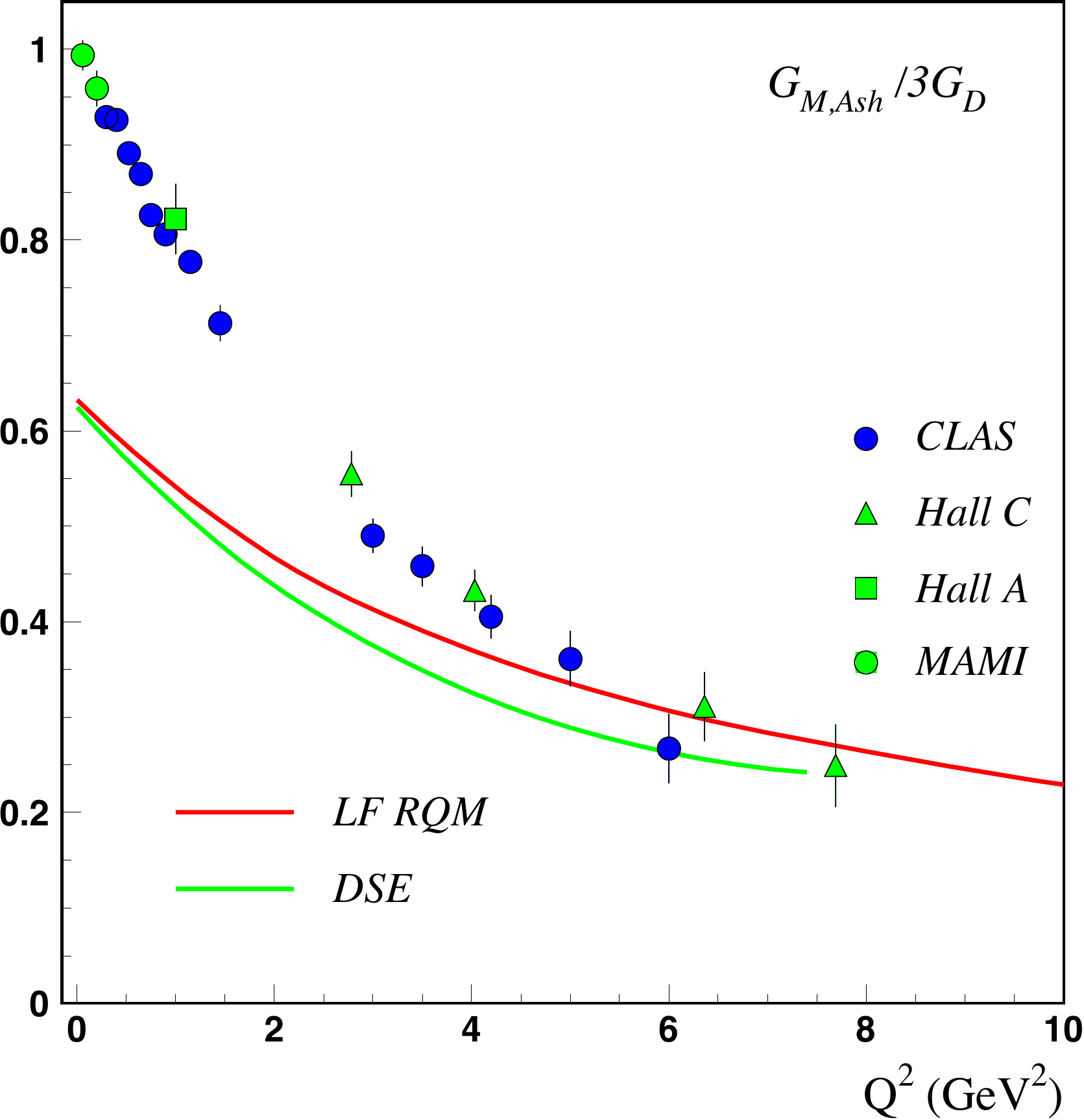}
&\hspace{+2mm}\vspace{+2mm}\includegraphics[width=0.28\textwidth,height=0.30\textwidth]{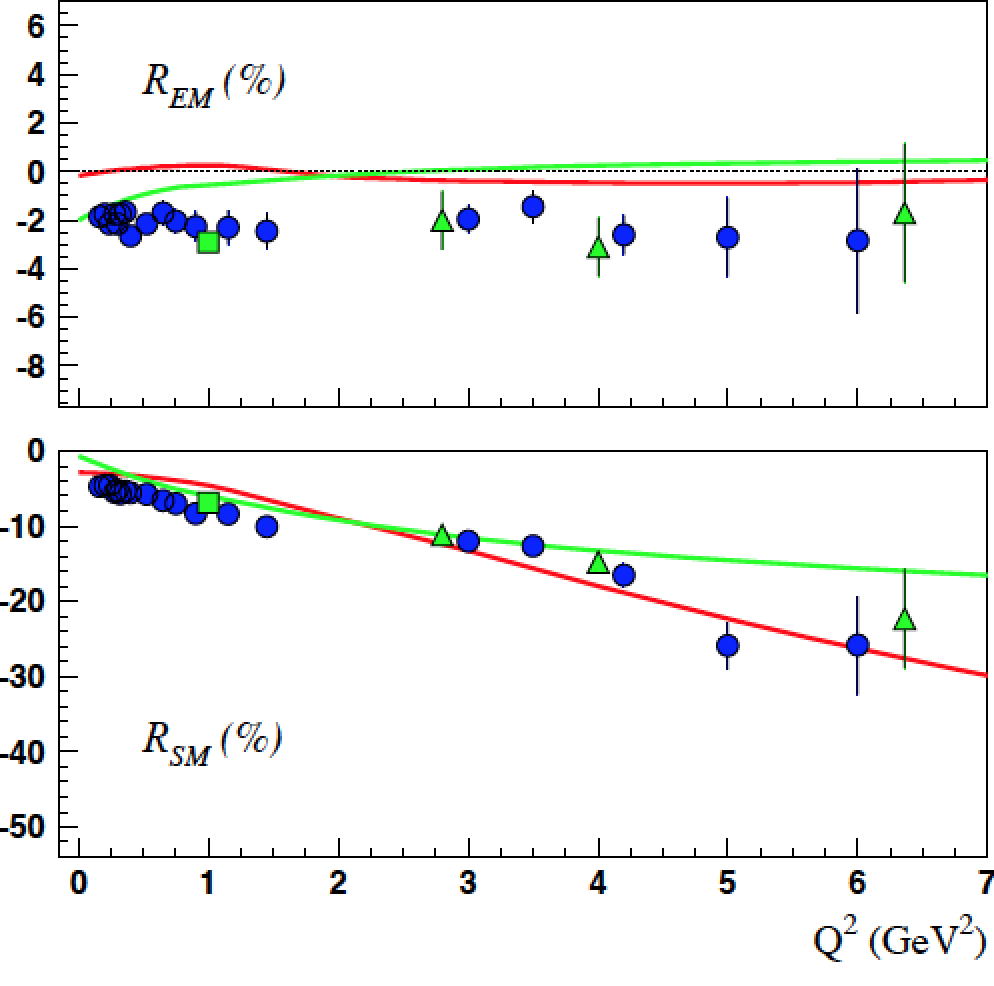}
&\hspace{+1mm}\includegraphics[width=0.30\textwidth,height=0.305\textwidth]{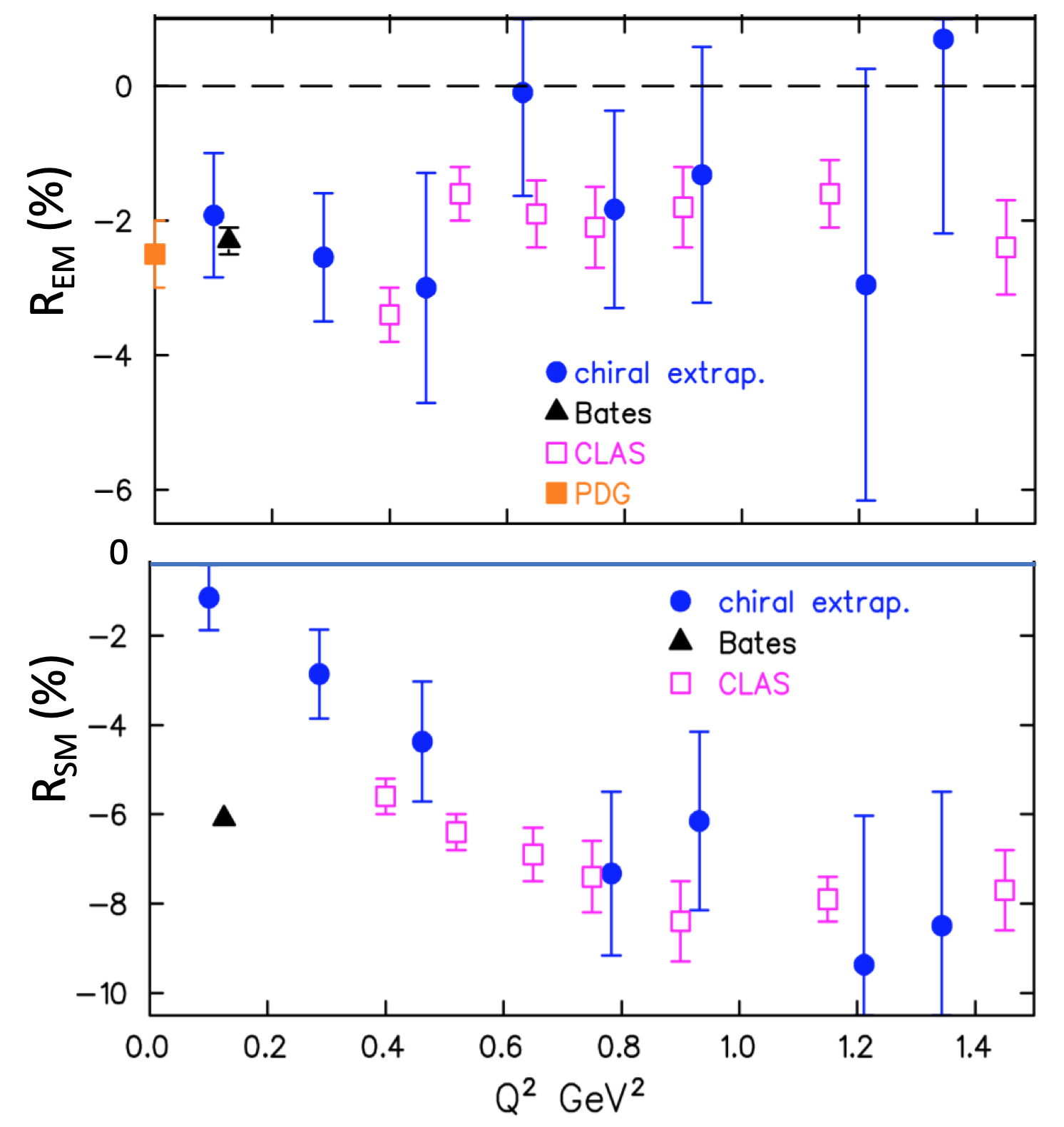}
\end{tabular}
\caption{The $N\Delta(1232)$ transition amplitudes. Left: The magnetic N$\Delta$ transition form factor normalized to the dipole form factor and compared with the Light-Front Relativistic Quark Model (LFRQM)\cite{Aznauryan:2012ec,Aznauryan:2016wwm} with running
quark mass, and with results using the Dyson-Schwinger Equation~\cite{Segovia:2014aza}. Both predictions are close to the data at high $Q^2$. At $Q^2 < 3$GeV$^2$ meson-baryon contributions are significant.  Middle: The electric (top) and scalar (bottom) quadrupole/magnetic-dipole ratios $R_{EM}$ and $R_{SM}$. Right:  $R_{EM}$ and $R_{SM}$ from Lattice QCD~\cite{Alexandrou:2007dt,Behrndt:2004km} compared to data in the low $Q^2$ domain.} 

\label{fig:Delta}
\end{figure*}
\begin{itemize}
    \item {\sl Dispersion Relations} have been employed in two ways: One is based on fixed-t dispersion relations 
    for the invariant amplitudes and was successfully used throughout the nucleon resonance region. Another way is 
   based on DR for the mulipole amplitudes of the $\Delta(1232)$ resonance, and allows getting functional forms
    of these amplitudes with one free parameter for each of them. It was employed for the analysis of 
    the more recent data.   
    \item {\sl The Unitary Isobar Model (UIM) } was developed in ~\cite{Drechsel:1998hk} from the effective Lagrangian 
    approach for pion photoproduction~\cite{Peccei:1969sb}. Background contributions from t-channel $\rho$ and $\omega$ 
    exchanges are introduced and the overall amplitude is unitarized in a K-matrix approximation. 
    \item {\sl Dynamical Models} have been developed, as SAID from pion
    photoproduction data~\cite{Arndt:2002xv}, the Sato-Lee model was developed in~\cite{Sato:1996gk}. Its essential feature 
    is the consistent description of $\pi N$ scattering and the pion electroproduction from nucleons. It 
    was utilized in the study of $\Delta(1232)$ excitations in the $ep\to ep\pi^0$ channel~\cite{Sato:2000jf}. The Dubna-Mainz-Taipei model~\cite{Kamalov:1999hs} builds unitarity via direct inclusion of the $\pi N$ 
    final state in the T-matrix of photo- and electroproduction. 

\end{itemize}  

\subsection{Light-quark baryon resonance electroproduction}
In order to learn from the meson electroproduction data about the internal spin and spatial electromagnetic structure, it is essential to have advanced models available with links to the fundamentals of $QCD$.    
\begin{figure}[hb!]
\resizebox{0.9\columnwidth}{!}{\includegraphics{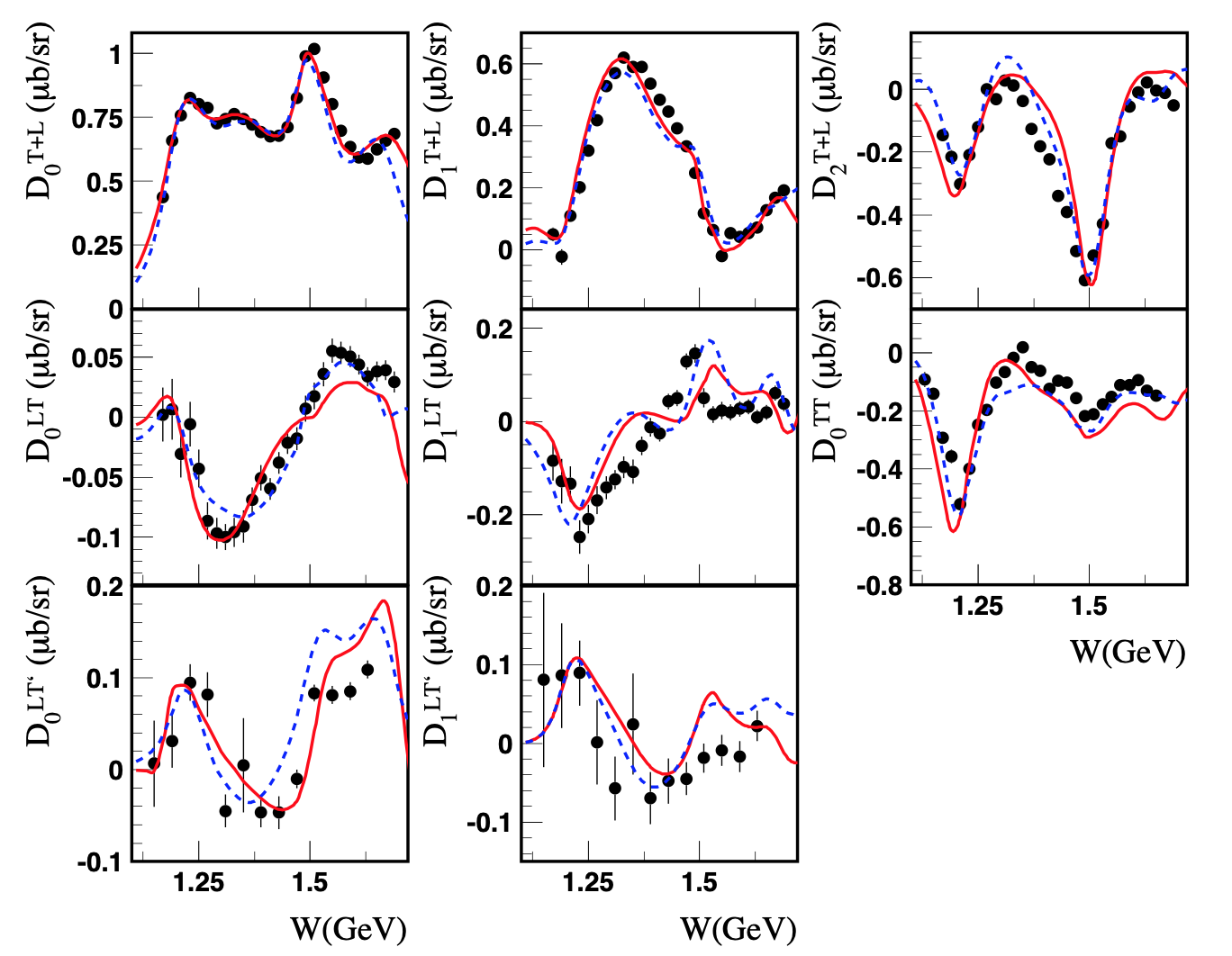}}
\caption{Sample of results of an analysis by the JLab group of the Legendre moments of $\vec{e}p \to e \pi^+ n$ structure functions in comparison with experimental data~\cite{CLAS:2007jpl} at $Q^2 = 2.44$~GeV$^2$. The solid (dashed) curves correspond to results obtained using the DR (UIM) approach.   }
\label{Hall-A}
\end{figure}

While most of the analyses have focused on single pseudoscalar meson production, such as $\gamma_vp \to N\pi, p\eta, K\Lambda$, $K\Sigma$, more recent work included the $p\pi^+\pi^-$ final state both in real photoproduction~\cite{CLAS:2018drk} as well as in electroproduction~\cite{Mokeev:2020hhu}. The 2-pion final state has more sensitivity to excited $N^*$ and $\Delta^*$ states in the mass range above 1.6~GeV, with several states dominantly coupling to $N\pi\pi$ final states, enabling the study of their electromagnetic transition form factors in the future.    

\subsubsection{The $N\Delta(1232)\frac{3}{2}^+$ transition }
The $\Delta^{++}$ isobar was first observed 70 years ago in Enrico Fermi's experiment that used a $\pi^+$ meson beam 
scattered off the protons in a hydrogen target~\cite{Anderson:1952nw}. The cross section showed a sharp rise above threshold towards a mass near 1200 MeV. While the energy of the 
pion beam was not high enough to see the maximum and the fall-off following the peak, a strong 
indication of the first baryon resonance was observed.  It took 12 more years and the development of the underlying symmetry in the quark model before a microscopic explanation of this observation could emerge. There was, however, a problem; while the existence of the $\Delta^{+,0,-}$ could be explained within the model, the existence of the $\Delta(1232)^{++}$, which within the quark model would correspond to a state $|u{\uparrow}u{\uparrow}u{\uparrow}\rangle$, was forbidden as it would have an 
overall symmetric wave function.   It took the introduction of para Fermi 
statistics~\cite{Greenberg:1964pe} what later became "color", to make the overall wave function anti-symmetric. 
In this way the $\Delta^{++}(1232)$ resonance may be considered a harbinger of the development of QCD.

The nucleon to $\Delta(1232)\frac{3}{2}^+$ transition is now well measured in a large range of $Q^2$~\cite{CLAS:2006ezq,CLAS:2001cbm,Frolov:1998pw}. At the real photon point, it is explained by a dominant magnetic transition from the  nucleon ground state to the $\Delta(1232)$ excited state. Additional contributions are related to small D-wave components in both the nucleon and the $\Delta(1232)$ wave functions leading to electric quadrupole and scalar quadrupole transitions. These remain in the few \% ranges at small $Q^2$. The magnetic transition is to $\approx 65\%$ given by a simple spin flip of one of the valence quarks as seen in Fig.~\ref{fig:Delta}. The remaining 35\% of the magnetic dipole strength is attributed to meson-baryon contributions. There exist extensive theoretical reviews of the $N\Delta(1232)$ transition in the lower $Q^2$ range are available~\cite{Pascalutsa:2005vq}, and more recent reviews that cover the full $Q^2$ covered by data~\cite{Aznauryan:2011qj,Tiator:2011pw}.    

\noindent The electric quadrupole ratio $R_{EM}$ was found as: 
\begin{eqnarray}
R_{EM}  \approx -0.02.
\end{eqnarray}  
There has been a longstanding prediction of asymptotic pQCD, that $R_{EM} \to +1$ at $Q^2 \to \infty$. Results on the magnetic transition 
form factor $\rm G_{Mn,Ash}$, defined in the Ash convention~\cite{Ash:1967rlw}, and on the quadrupole transition ratios are shown in Fig.~\ref{fig:Delta}.    
\begin{figure*}[th!]
\centering
\begin{tabular}{ccc}
\includegraphics[width=0.32\textwidth,height=0.32\textwidth]
{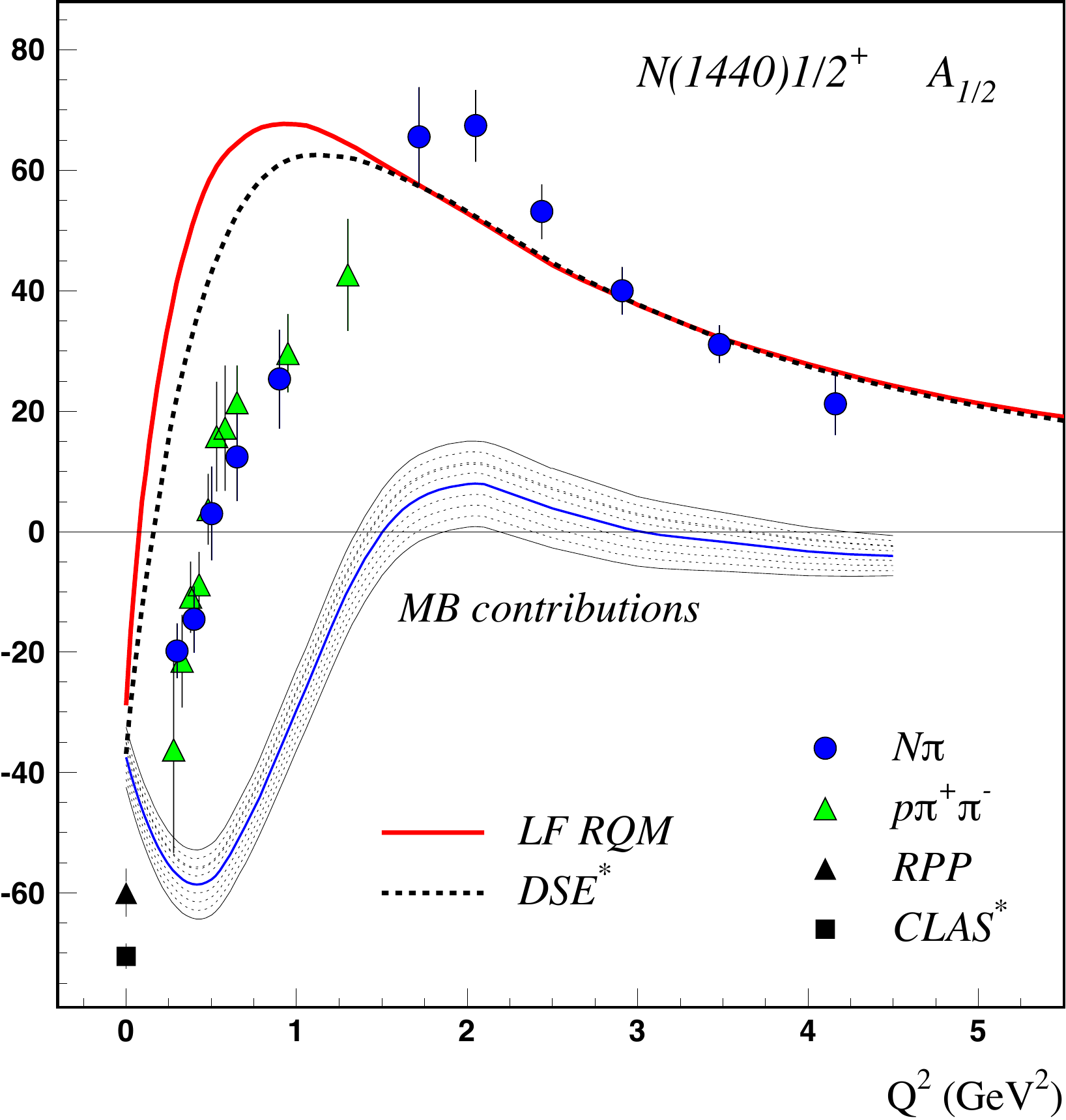}
\hspace{1mm}
\includegraphics[width=0.32\textwidth,height=0.32\textwidth]
{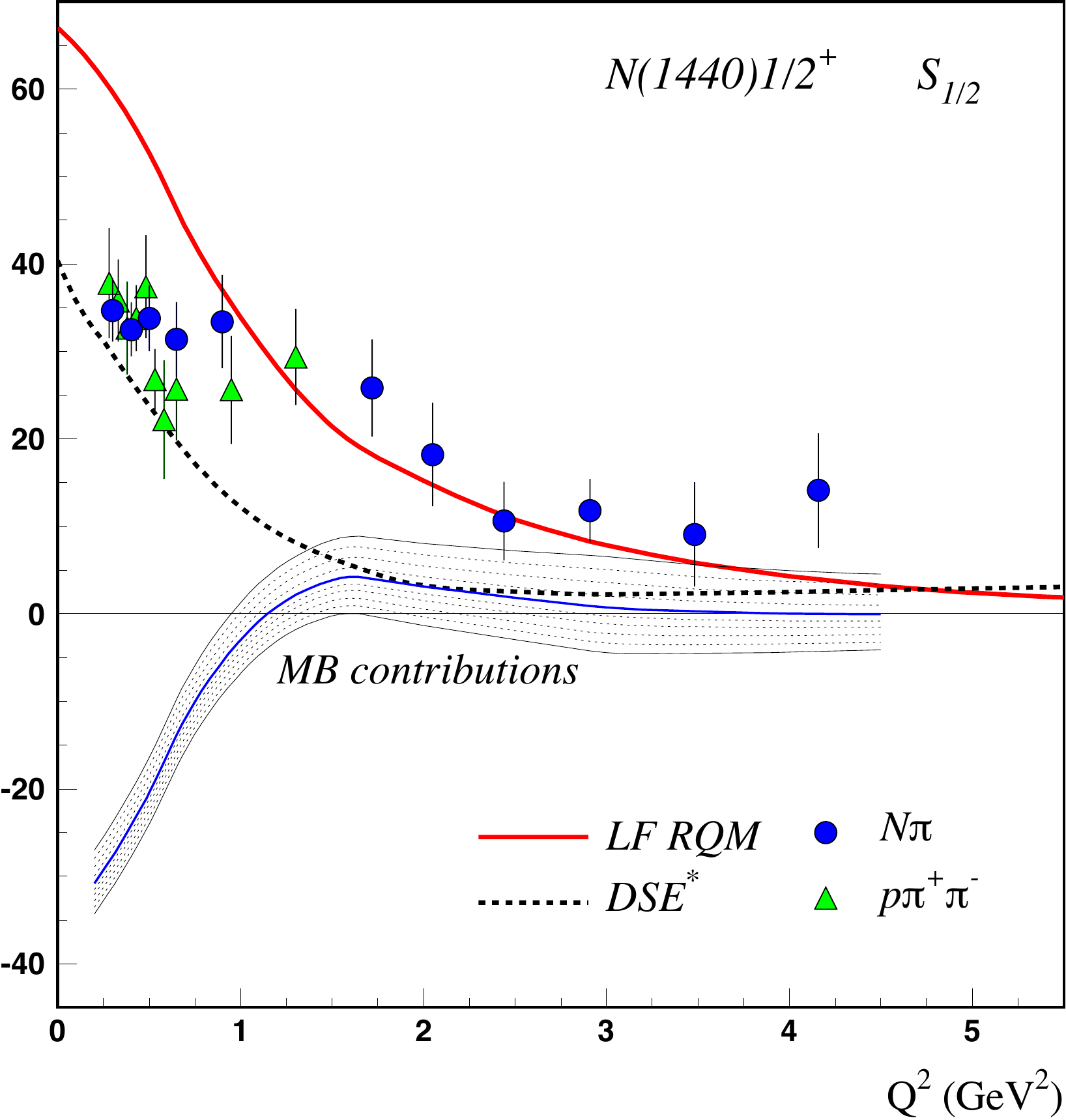}
\vspace{1.5mm}
\includegraphics[width=0.34\textwidth,height=0.34\textwidth]
{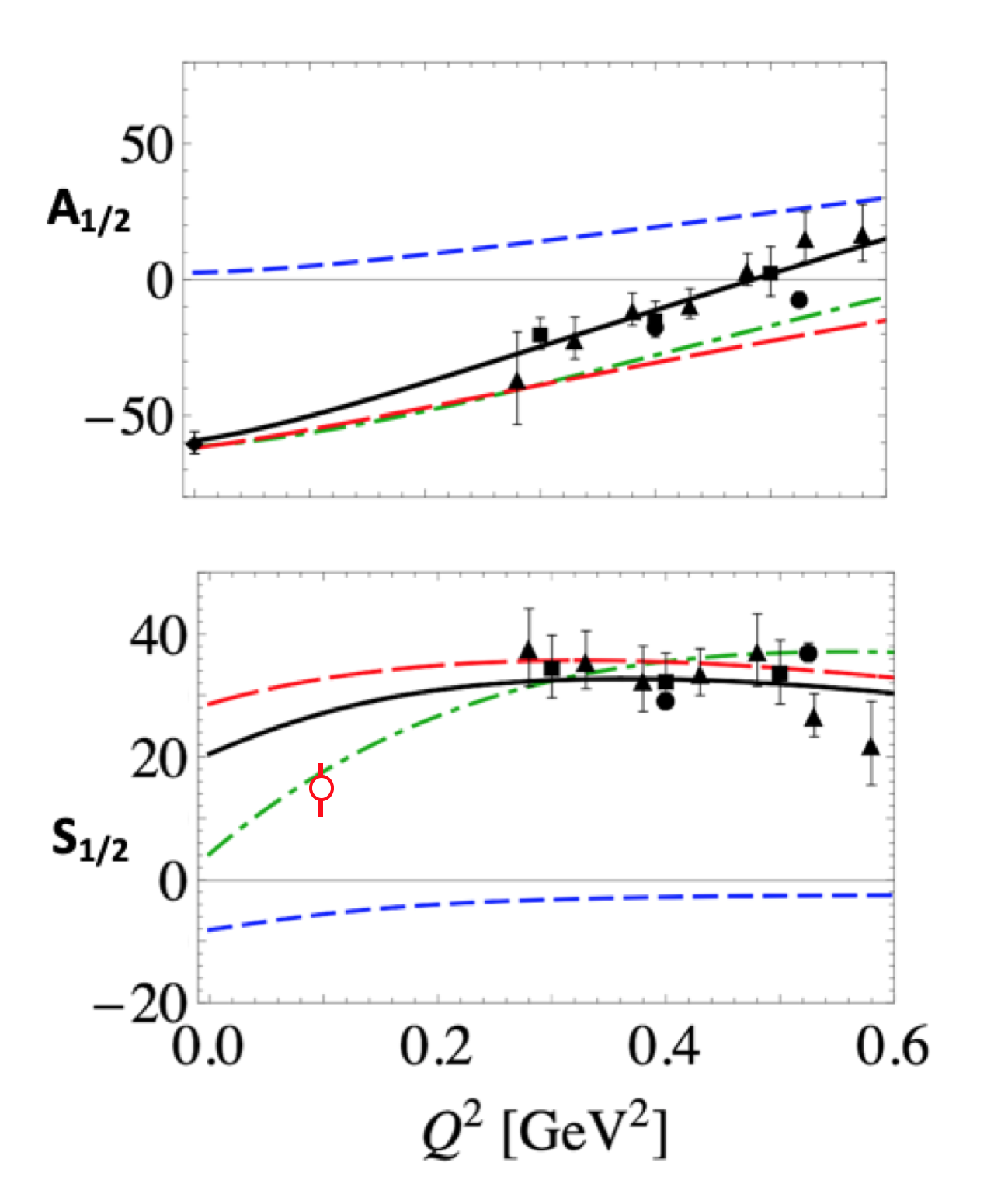}
\end{tabular}
\caption{Helicity transition amplitudes for the proton to Roper $N(1440)\frac{1}{2}^+$ excitation in units of $10^{-3}\rm{GeV}^{-1/2}$ compared to various model calculations; see text. Left: Transverse $A_{1/2}$ amplitude. Middle: Scalar $S_{1/2}$ amplitude.  Right: Helicity amplitudes of the Roper resonance at low $Q^2$. Data are compared to calculations within Effective Field Theory~\cite{Bauer:2014cqa} shown in solid black lines. The other broken lines are parts of the full calculations. The data are from~\cite{CLAS:2009ces,CLAS:2012wxw,Drechsel:2007if}. The open red circle at $Q^2 \approx 0.1$GeV$^2$ is the result of an analysis of $ep \to ep\pi^0$ data from \cite{Stajner:2017fmh}. }
\label{p11}
\end{figure*}
$\rm G_{Mn,Ash}$ is shown normalized to the dipole form factor, but indicates a much faster $Q^2$ fall-off compared to that. 
In comparison to the advanced LFRQM with momentum-dependent constituent quark mass, and with the Dyson-Schwinger Equation (DSE-QCD) results, there is 
good agreement at the high-$Q^2$ end of the data. The discrepancy at small $Q^2 = 0$ is likely due to the meson-baryon contributions at small $Q^2$, which are not modeled in either of the calculations.     

The quadrupole ratio $R_{EM}$ shows no sign of departing significantly from its value at $Q^2=0$, even at the highest $Q^2 \approx 6.5$~GeV$^2$. 
Both calculations barely depart from $R_{EM}=0$, and remain near zero at all $Q^2 > 2$~GeV$^2$. This indicates that the negative constant value shown by the 
data is likely due to meson-baryon contributions that are not included in the theoretical models.   
For the scalar quadrupole ratio $R_{SM}$ the asymptotic prediction in holographic QCD (hQCD) ~\cite{Grigoryan:2009pp} is: 
\begin{eqnarray}
R_{SM} = \frac{Im S_{1+}}{Im M_{1+}} \to -1, ~~{\rm at ~Q^2 \to \infty},  
\end{eqnarray}  
while $R_{EM}$ in hQCD is predicted to approach +1 asymptotically. 
The $R_{SM}$ data show indeed a strong trend towards increasing negative values at larger  $Q^2$, semi-quantitatively described by both calculations at $Q^2<4$~GeV$^2$. The Dyson-Schwinger equation approach predicts a flattening of $R_{SM}$ at $Q^2>4$~GeV$^2$, while the Light Front relativistic Quark Model predicts a near constant negative slope of $R_{SM}(Q^2)$ also at higher $Q^2$.

\subsubsection{The Roper resonance $N(1440)\frac{1}{2}^+$}

The Roper resonance, discovered in 1964~\cite{Roper:1964zza} in a phase shift analysis of elastic $\pi N$ scattering data, has been differently interpreted for half a century. In the non-relativistic quark model (nrQM), the state is the first radial excitation of the nucleon ground state with a mass 
expected around 1750 MeV, much higher than the measured Breit-Wigner mass of $\approx 1440$~MeV. This discrepancy is now understood as the consequence  of a dynamical coupled channel effect that shifts the mass below the mass of the $N(1535){1/2}^-$ state, the negative-parity partner of the nucleon~\cite{Suzuki:2009nj}. Another problem with the quark model was the sign of the transition form factor $A_{1/2}(Q^2=0)$, predicted in the nrQM as large and positive, while experimental analyses showed a negative value.

 These discrepancies resulted in different interpretations of the state that could only be resolved with electroproduction data from CLAS at Jefferson Lab, the development of continuous QCD approximations in the Dyson-Schwinger equation approach~\cite{Segovia:2015hra} and Light Front Relativistic QM with momentum-dependent quark masses~\cite{Aznauryan:2012ec} shown in Fig.~\ref{p11}, and Lattice data~\cite{Mathur:2003zf,Lin:2011da}. A recent review of the history and current status of the Roper resonance, is presented in a colloquium-style article published in Review of Modern Physics~\cite{Burkert:2017djo}. 
 
Descriptions of the baryon resonance transitions form factors, including the Roper resonance $N(1440)\frac{1}{2}^+$, have also been carried out within holographic models~\cite{deTeramond:2011qp,Ramalho:2017pyc}. In the range $Q^2 < 0.6$~GeV$^2$, calculations based on meson-baryon degrees of freedom and effective field theory~\cite{Bauer:2014cqa} have been successfully performed, as may be seen in Fig.~\ref{p11}. Earlier model descriptions, such as the Isgur-Karl model that describe the nucleon as a system of 3 constituent quarks in a confining potential and a one-gluon exchange contribution leading to a magnetic hyperfine splitting of states~\cite{Isgur:1978xj,Isgur:1977ef}, and the relativized version of Capstick~\cite{Capstick:1986ter} have popularized the model that became the basis for many further developments and variations, e.g. the light front relativistic quark model, and the hypercentral quark model~\cite{Giannini:2003xx}. Other models were developed in parallel. The cloudy bag  model~\cite{Thomas:1981vc} describes the nucleon as a bag of 3 constituent quarks surrounded by a cloud of pions. It has been mostly applied to nucleon resonance excitations in real photoproduction,  $Q^2=0$~\cite{Bermuth:1988ms,Thomas:1981vc}, with some success in the description of the $\Delta(1232)\frac{3}{2}^+$ and the Roper resonance transitions. 

\begin{figure}[th!]
\centering
\begin{tabular}{cc}
\hspace{-2mm}\includegraphics[width=0.48\textwidth,height=0.25\textwidth]{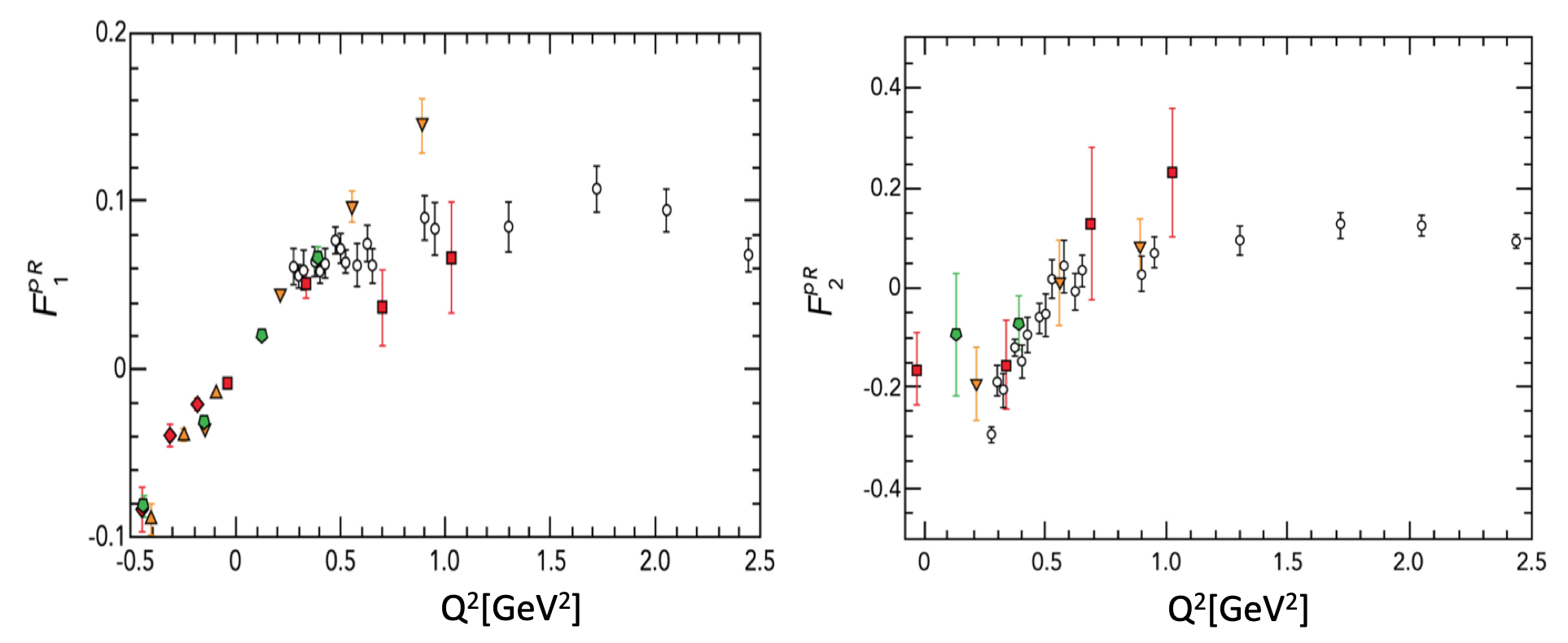}
\end{tabular}
\caption{Dirac and Pauli transition form factors $F_1$ and $F_2$ for the proton to $N(1440)1/2^+$  transition compared to Lattice QCD calculations~\cite{Lin:2011da} 
with pion masses (in GeV): 0.39 (red squares), 0.45 (orange triangles), and 0.875 
(green circles) on the $N_f = 2 + 1$ anisotropic lattices, compared to CLAS results (black circles). The $F_1$ and $F_2$ form factors are linear combinations of the $A_{1/2}$ and $S_{1/2}$ amplitudes. }
\label{fig:Roper-Lattice}
\end{figure}

There is agreement with the data at $Q^2 > 1.5$~GeV$^2$ for these two states, while the meson-baryon contributions for 
the $\Delta(1232)$ are more extended, and agreement with the quark based calculations is reached at $Q^2 > 4$~GeV$^2$.  
The calculations deviate significantly from the data 
at lower $Q^2$, which indicates the presence of non-quark core effects.  For the Roper resonance such contributions have been described 
successfully in dynamical meson-baryon models~\cite{Obukhovsky:2011sc} and in effective field theory~\cite{Bauer:2014cqa}. 
Calculations on the Lattice for the N-Roper transition form factors $F_1^{pR}$ and $F_2^{pR}$, which are combinations of the 
transition amplitudes $A_{1/2}$ and $S_{1/2}$, have been carried out with dynamical quarks~\cite{Lin:2011da}. 
The results agree well with the data in the range $Q^2 < 1.0$~GeV$^2$, where data and calculations overlap~Fig.~\ref{fig:Roper-Lattice}.

New electroproduction data on the Roper~\cite{Stajner:2017fmh} and on several higher mass states have been obtained in the 2-pion channel, specifically in $ep\to e'p\pi^+\pi^-$~\cite{Mokeev:2015lda}.
The mass of the Roper state has been computed on the Lattice and extrapolated to the physical pion mass, showing good agreement with the physical value measured with a Breit-Wigner parametrization. It should be noted that the Roper mass measured at the pole in the complex plane is significantly different from the value obtained using the BW ansatz.   
\begin{figure*}[ht!]
\centering
\includegraphics[width=0.95\textwidth,height=0.3\textwidth]{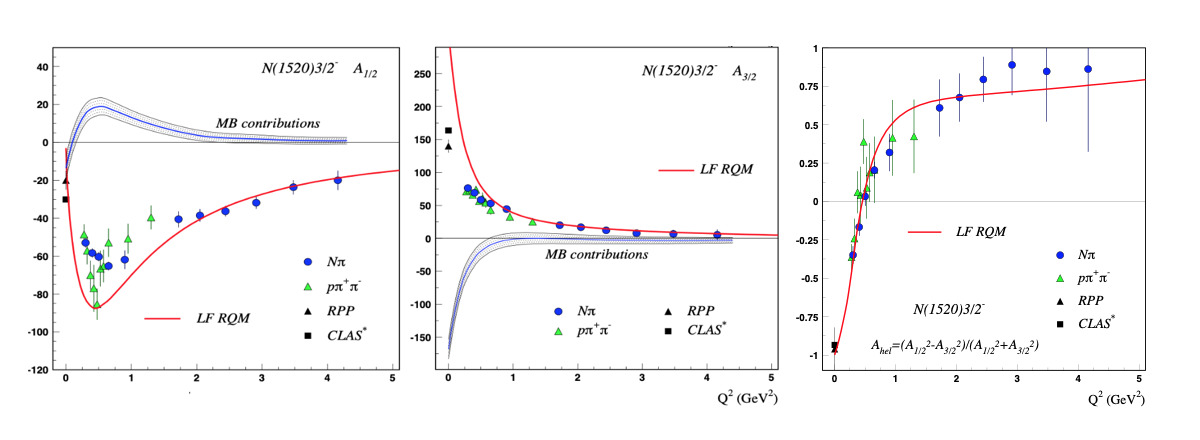}
\caption{The transverse helicity transition amplitudes of $N(1520)\frac{3}{2}^-$ versus $Q^2$, compared to the LFRQM, $A_{1/2}$ (left), $A_{3/2}$ (middle).  The shaded area indicates the
contribution from non-quark contributions as estimated from the difference of the measured data points and the LF RQM contribution, likely due to hadronic contributions. Right: Helicity asymmetry $A_{hel}$, as defined in Eq.~\ref{Ahel}. Graphics from Ref.~\cite{Aznauryan:2011qj}}
\label{fig:N1520}
\end{figure*}
Supported by an extensive amount of single pion electroproduction data, covering the full phase space in the pion polar and azimuthal center-of-mass angles, and accompanied by several theoretical modeling, we can summarize our current understanding of the $N(1440)\frac{1}{2}^+$ state as follows:  
\begin{itemize}
    \item The Roper resonance is, at heart, the first radial excitation of the nucleon.
    \item It consists of a well-defined dressed-quark core, which plays a role in determining the system's properties at all length scales, but exerts a dominant influence on probes with $Q^2 > m_N^2$, where $m_N$ is the nucleon mass.
    \item The core is augmented by a meson cloud, which both reduces the Roper's core mass by $\approx 20\%$, thereby solving the mass problem that was such a puzzle in constituent quark model treatments, and, at low $Q^2$, contributes an amount to the electroproduction transition form factors that is comparable in magnitude with that of the dressed quark core, but vanishes rapidly as $Q^2$ is increased beyond $m_N^2$. 
\end{itemize}
As stated in the conclusions of ~\cite{Burkert:2017djo}: "The fifty years of 
experience with the Roper resonance have delivered lessons that 
cannot be emphasized too strongly. Namely, in attempting to predict and explain the QCD spectrum, one must fully consider the impact of meson-baryon final state interactions and the coupling between channels and states that they generate, and look beyond merely locating the poles in the S-matrix, which themselves reveal little structural information, to also consider the $Q^2$ dependencies of the residues, which serve as a penetrating scale-dependent probe of resonance composition."

\subsubsection{The helicity structure of the $N(1520)\frac{3}{2}^-$ }
The $N(1520)\frac{3}{2}^-$ state corresponds to the lowest excited nucleon resonance with $J^P=\frac{3}{2}^-$. Its helicity structure is defined by the $Q^2$ dependence of the two transverse transition amplitudes $A_{1/2}$ and $A_{3/2}$.  They are both shown in Fig.~\ref{fig:N1520}. A particularly interesting feature of this state is that at the real photon point, $A_{3/2}$ is strongly dominant, while $A_{1/2}$ is very small. However, at high $Q^2$, $A_{1/2}$ is becoming dominant, while $A_{3/2}$ drops rapidly. This behavior is qualitatively consistent with the expectation of asymptotic QCD, which predicts the transition helicity amplitudes to behave like:      
\begin{eqnarray} 
\hspace{2.5cm} A_{1/2} \propto \frac{a}{Q^3},  A_{3/2} \propto \frac{b}{Q^5}\,.
\end{eqnarray} 
The helicity asymmetry 
\begin{eqnarray}
\hspace{2.5cm}A_{hel} = \frac{A_{1/2}^2 - A_{3/2}^2}{ A_{1/2}^2 + A_{3/2}^2}, \label{Ahel}
\end{eqnarray} 
shown in Fig.~\ref{fig:N1520}, illustrates this rapid change in the helicity structure of the $\gamma_v pN(1520){3/2}^-$ transition. At $Q^2 > 2~$GeV$^2$, $A_{1/2}$ fully dominates the process.  

\begin{figure}[bh!] 
\includegraphics[width=0.4\textwidth,height=0.35\textwidth]{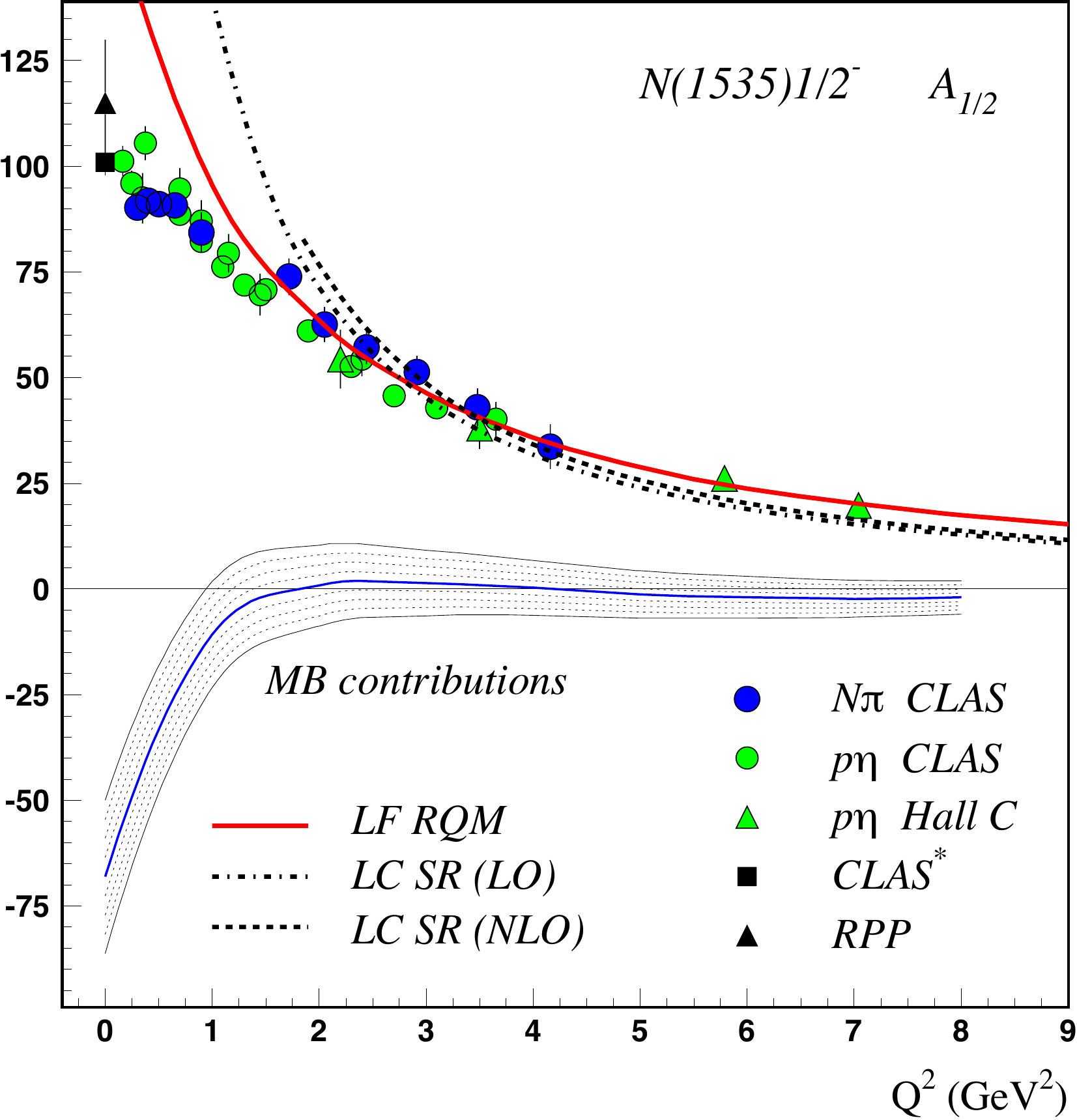}
\caption{Right: The transverse transition helicity amplitude $A_{1/2}$ versus $Q^2$. At $Q^2 >2$GeV$^2$ the data are well described by the light-cone sum rules LCSR~\cite{Braun:2009jy}. The light front relativistic quark model (LFRQM)~\cite{Aznauryan:2017nkz} describes that data at $Q^2 > 1$GeV$^2$.}
\label{N1535}
\end{figure}
\subsubsection{Transition Form Factors of  $N(1535)\frac{1}{2}^-$ - A state with a hard quark core. }   
This state is the parity partner state to the ground state nucleon, with the same spin 1/2 but 
with opposite parity, its quark content requires an orbital L=1 excitation in the transition 
from the proton. In the $SU(6)\otimes O(3)$ symmetry scheme, the state is a member of the 
$[70,1^-]$ super multiplet. This state couples equally to $N\pi$ and to $N\eta$ final state. 
It has therefore be probed using both decay channels $ep \to e p \eta$ and $ep \to eN\pi^{+,0}$. 
Because of isospin $I=1/2$ for nucleon states, the coupling to the charged $\pi^+n$ channel 
is preferred over $\pi^0p$ owing to the Clebsch-Gordon coefficients. 

 The $A_{1/2}$ helicity amplitude for the $\gamma pN(1535)\frac{1}{2}^-$ resonance excitation  
shown in Fig.~\ref{N1535} represents the largest range in $Q^2$ of all nucleon states for which resonance transition form factors have been measured as part of the broad experimental program at JLab.  

For this state, as well as for the $N(1440)\frac{1}{2}^+$ state, advanced relativistic quark model calculations~\cite{Aznauryan:2015zta}, DSE-QCD calculations~\cite{Segovia:2015hra} and 
Light Cone sum rule results~\cite{Anikin:2015ita} are available, employing QCD-based modeling of the excitation of the quark core for the first time.
\begin{figure*}[ht!]
\centering
\begin{tabular}{ccc}
{\includegraphics[width=0.30\textwidth,height=0.30\textwidth]{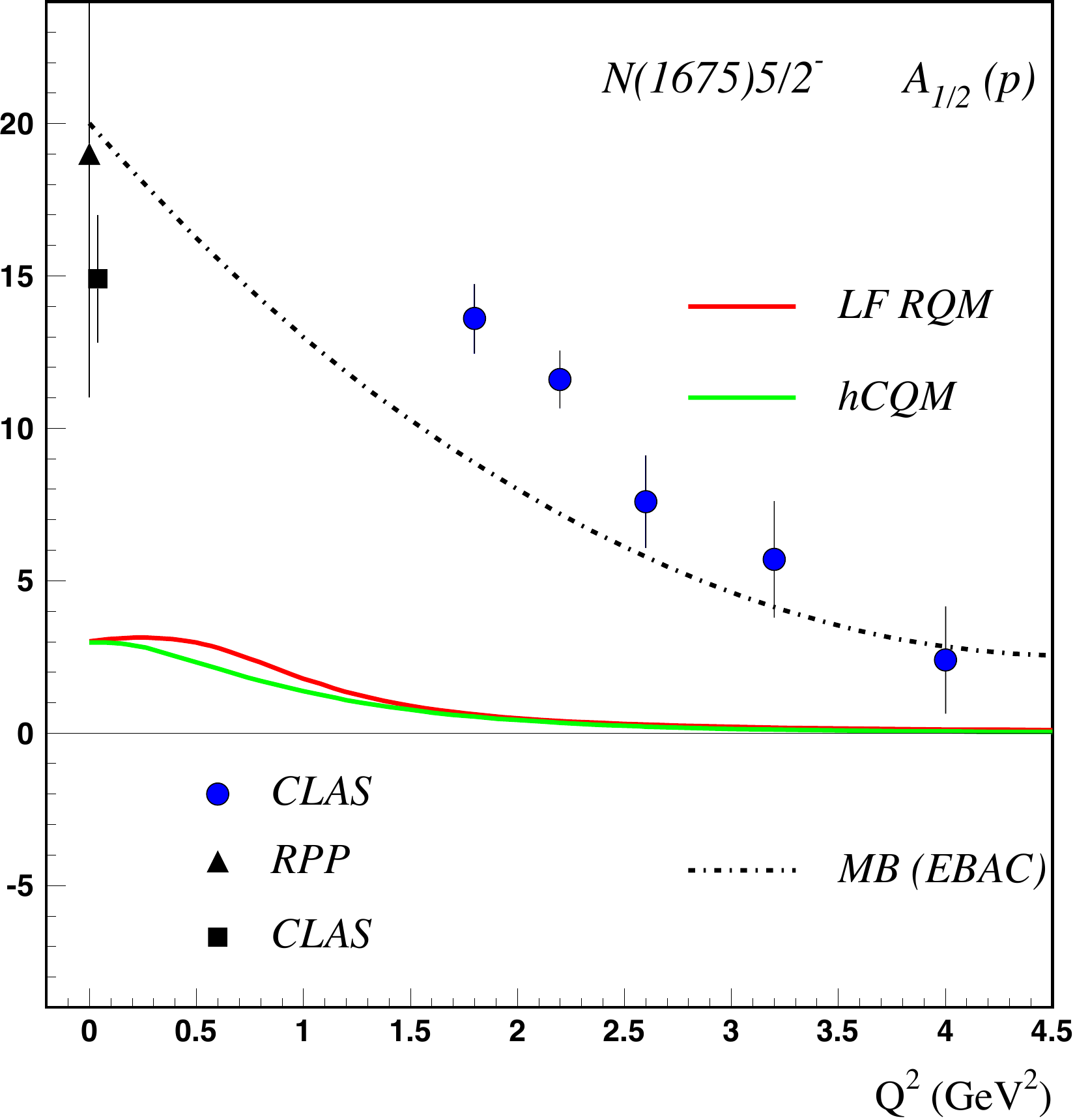}}
{\includegraphics[width=0.30\textwidth,height=0.30\textwidth]{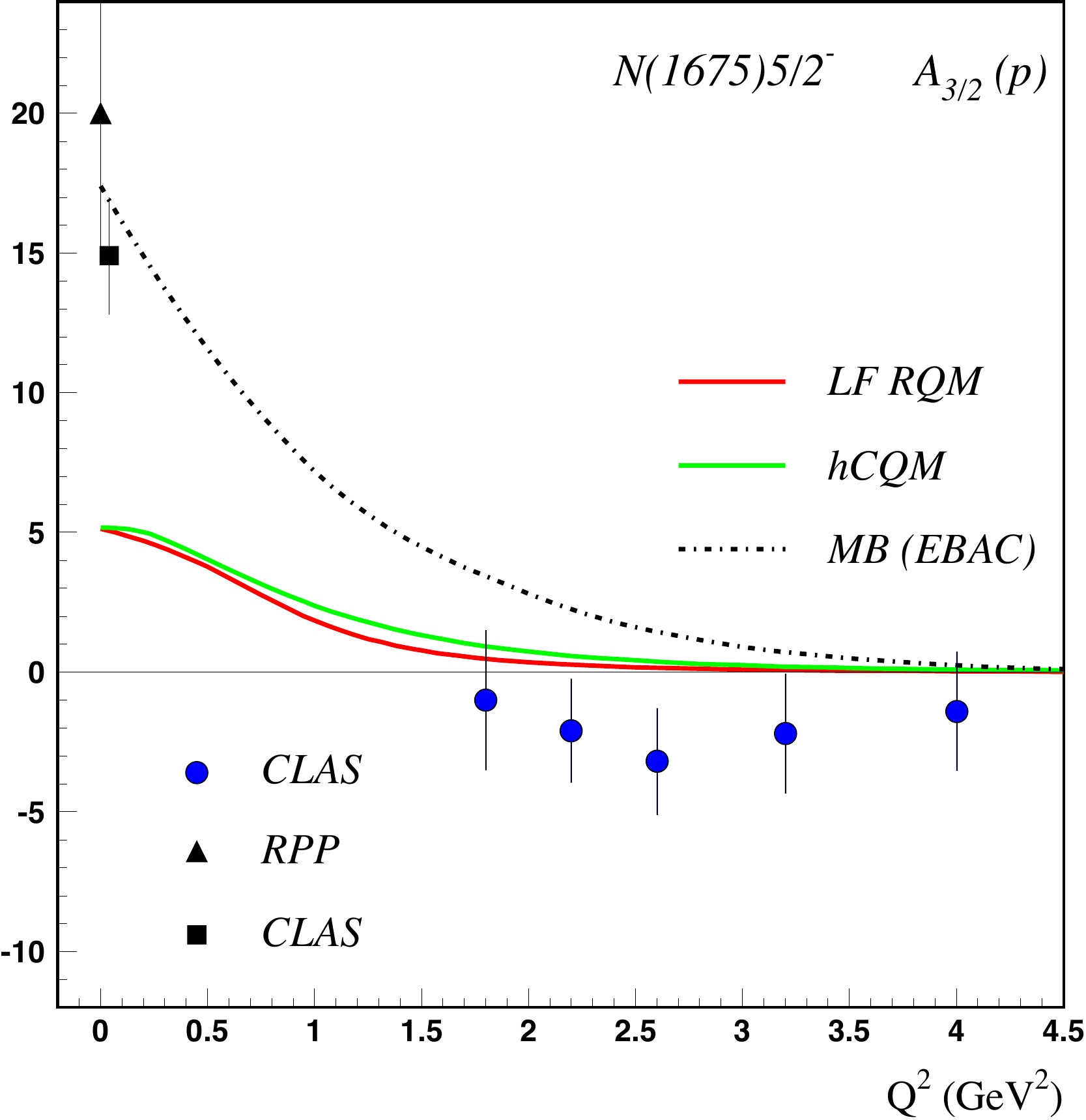}}
{\includegraphics[width=0.32\textwidth,height=0.31\textwidth]{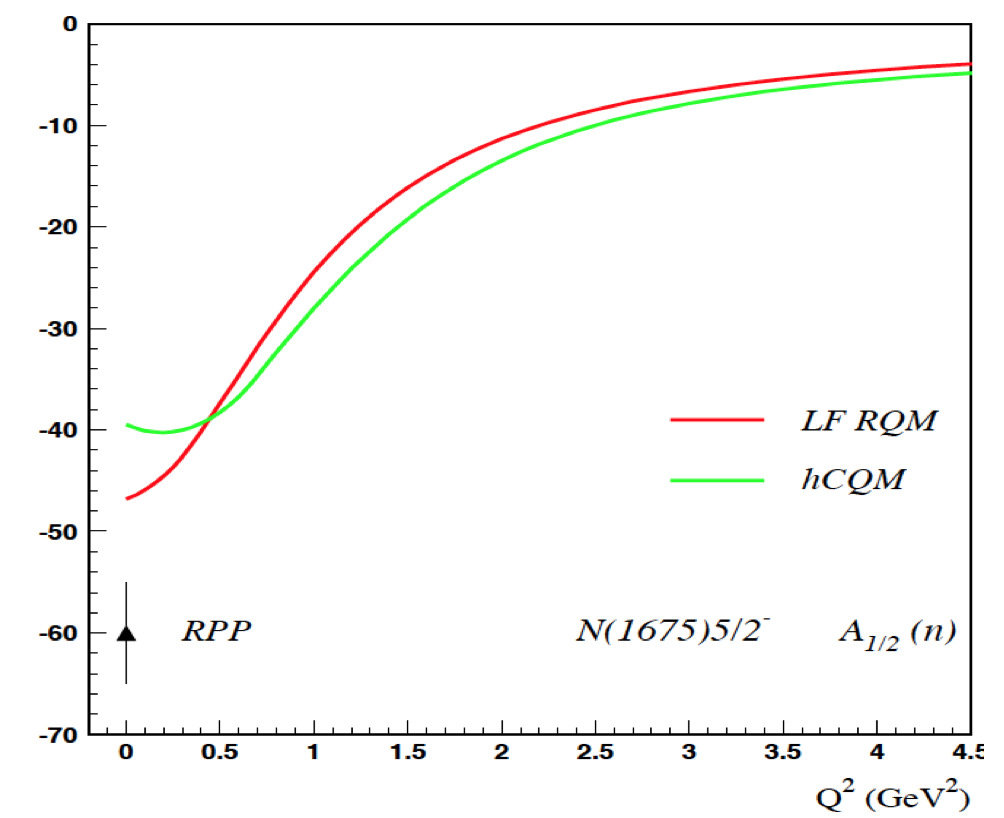}}
\end{tabular} 
\caption{The transverse amplitudes of the proton to $N(1675)\frac{5}{2}^-$ transition compared to the LF RQM~\cite{Aznauryan:2014xea}, hypercentral QM~\cite{Santopinto:2012nq}, and contributions from meson-baryon (MB) coupled channel dynamics~\cite{Julia-Diaz:2007mae}. Left: $A_{1/2}$, Middle:$A_{3/2}$. Both quark models predict very small amplitudes for the proton, while the meson-baryon contributions estimate is large and is close to the data. Right: $A_{1/2}$ for neutron target (only photoproduction data available) compared to the LFRQM and hCQM. Both quark models predict large amplitudes for neutrons, more than factor 10 compared to protons at $Q^2=0$. Assuming similar meson-baryon contributions as in the proton case with opposite sign could quantitatively explain the single measured value at the photon point. }
\label{N1535-N1675}
\end{figure*}

The transverse transition amplitude $A_{1/2}$ of $N(1535)\frac{1}{2}^{-}$ is a prime example of the power of meson electroproduction to unravel the internal structure of the resonance transition. In another section of the Volume "50 Years of Quantum Chromodynamics" ~\cite{Burkert:2022hjz}, the nature of this state is discussed as a possible example of a dynamically generated resonance. The electroproduction data shown here reveal structural aspects of the state and its nature that require a different interpretation. The transition form factor $A_{1/2}$ of the state, shown in Fig.~\ref{N1535}, is quantitatively reproduced over a large range in $Q^2$ by two alternative approaches, the LFRQM and the LCSR. Both calculations are based on the assumptions of the presence of a 3-quark core. Notice that there is deviation from the quark calculations at $Q^2 <1-2$~GeV$^2$, highlighted as the shaded area in Fig.~\ref{N1535}, which may be assigned to the presence of non-quark contributions. Attempts to compute the transition form factors within strictly dynamical models have not succeeded in explaining the available data~\cite{Jido:2007sm}. The discrepancy could be resolved if the character of the probe, meson (pion) in the case of hadron interaction and short range photon interaction in the case of electroproduction, probe different parts of the resonance's spatial structure: peripheral in case of meson scattering and short distance behavior in electro-production. The peripheral meson scattering and low $Q^2$ meson photo-production reveal the dynamical features of the state, whereas high $Q^2$ electroproduction reveals the structure of the quark core.

\subsubsection{The $N(1675)\frac{5}{2}^-$ resonance reveals the meson-baryon contributions}

In previous discussions we have concluded that meson-baryon degrees of freedom provide significant strength to the 
resonance excitation in the low-$Q^2$ domain where quark based approaches LF RQM, DSE/QCD, and LC SR calculations fail to 
reproduce the transition amplitudes quantitatively. Our conclusion rests, in part, with this assumption. 
But, how can we be certain of the validity of this conclusion?

The $N(1675)\frac{5}{2}^-$ resonance allows us to test this assumption, quantitatively. Figure~\ref{N1535-N1675} shows 
our current knowledge of the transverse helicity amplitudes $A_{1/2}(Q^2)$ and $A_{3/2}(Q^2)$, for proton target compared to  
RQM~\cite{Aznauryan:2017nkz} and 
hypercentral CQM~\cite{Santopinto:2012nq} calculations. The specific quark transition for a $J^P = \frac{5}{2}^-$ state belonging to the 
$[SU(6)\otimes O(3)] = [70, 1^-]$ supermultiplet configuration, in non-relativistic approximation prohibits the transition from the
 proton in a single quark transition.
This suppression, known as the Moorhouse selection rule~\cite{Moorhouse:1966jn}, is valid for the transverse transition 
amplitudes $A_{1/2}$ and $A_{3/2}$ at all $Q^2$. It should be noted that this selection rule does apply to the 
transition from a proton target, it does not apply to the transition from the neutron, which is consistent with the data.  Modern quark models that go beyond 
single quark transitions, confirm quantitatively the suppression resulting in very 
small amplitudes from protons but large ones from neutrons. Furthermore, a direct computation of the hadronic 
contribution to the transition from protons confirms this (Fig.~\ref{N1535-N1675}).  The measured helicity amplitudes off the protons are almost  
exclusively due to meson-baryon contributions as the dynamical coupled channel (DCC) calculation indicates (dashed line). 
The close correlation of the 
DCC calculation and the measured data for the case when quark contributions are nearly absent,   
supports the phenomenological description of the helicity amplitudes in terms of a 3-quark core that dominate at high $Q^2$ and 
meson-baryon contributions that can make important contributions at lower $Q^2$.

\subsubsection{Resonance lightfront transition charge densities.} 

Knowledge of the helicity amplitudes in a large $Q^2$ allows for the determination of the transition charge densities on the light cone in transverse impact parameter space ($b_x, b_y$)~\cite{Tiator:2008kd}. 
The relations between the helicity transition amplitudes and the Dirac and Pauli resonance transition form factors are given by: 
\begin{eqnarray}
A_{1/2} = e \frac{Q_-}{\sqrt{K}(4M_NM^*)^{1/2}}\{F_1^{NN^*} + F_2^{NN^*}\}~~~~~ \\
S_{1/2} = e \frac{Q_-}{\sqrt{K}(4M_NM^*)^{1/2}}\left (\frac{Q_+Q_-}{2M^*}\right) \frac{(M^*+M_N)}{Q^2} \nonumber \\
\times \{F_1^{NN^*} - \frac{Q^2}{(M^*+M_N)^2}F_2^{NN^*}\}\,,~~~~~
\end{eqnarray}

\noindent
where $M^*$ is the mass of the excited state $N^*$, $K=\frac{{M^*}^2 - M^2_N}{2M^*}$ is the equivalent photon energy, $Q_+$ and $Q_-$ are short hands for $Q_\pm \equiv \sqrt{M^* \pm M_N)^2 + Q^2}$. The charge and magnetic lightfront transition densities $\rho_0^{NN^*}$ and $\rho_T^{NN^*}$, respectively, are given as:
\begin{eqnarray}
\rho_0^{NN^*}(\vec{b}) = \int_0^{\infty}\frac{dQ}{2\pi}J_0(bQ)F_1^{NN^*}(Q^2) \\
\rho_T^{NN^*}(\vec{b}) = \rho_0^{NN^*}(\vec{b}) + \sin(\phi_b - \phi_s)\times \nonumber\\
\int_0^{\infty}\frac{dQ}{2\pi}\frac{Q^2}{(M^*+M_N)}J_1(bQ)F_2^{NN^*}(Q^2)\,.
\end{eqnarray}
\begin{figure}[thb!]
\centerline{\resizebox{0.95\columnwidth}{!}{\includegraphics{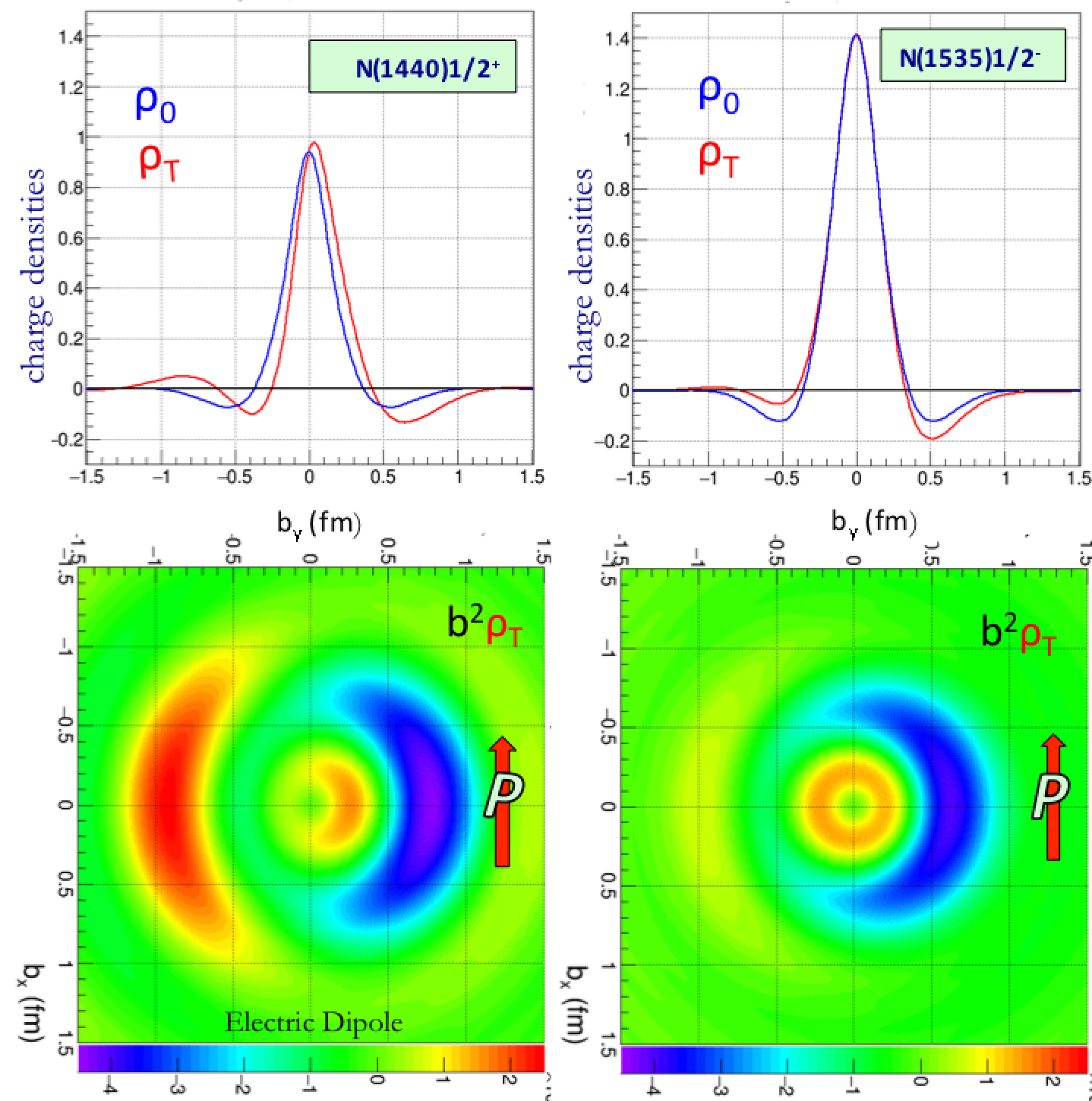}}}
\caption{Left panels: $N(1440)$, top: projection of charge densities on $b_y$, bottom: transition charge densities when the proton  is spin polarized along $b_x$. Right panels: same for $N(1535)$. Note that the densities are scaled with 
$b^2$ to emphasize the outer wings. Color code:negative charge is blue, positive charge is red. Note that all scales
are the same for ease of comparison~\cite{Burkert:2018oyl}. Graphics credit: F.X. Girod.}
\label{charge_densities}
\end{figure}

\begin{figure}[tph!]
\centerline{\resizebox{0.8\columnwidth}{!}{\includegraphics{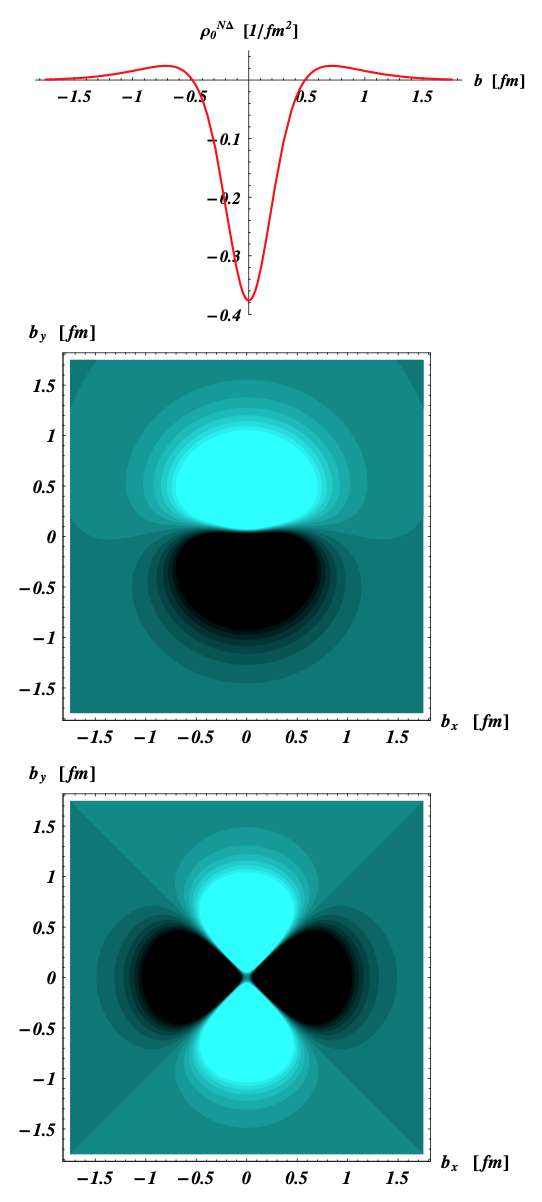}}}
\caption{Quark transverse transition charge density corresponding to the $p \to \Delta^+$ transition. Light color indicates positive charge, dark color indicates negative charge. Top: p and $\Delta$ are unpolarized. Middle: p and $\Delta$ are polarized along $b_x$ axis generating an electric dipole along the $b_y$ axis. Bottom: Quadrupole contribution to transition density. Graphics adapted from~\cite{Carlson:2007xd}. } 
\label{p-Delta-TCD}
\end{figure}

Similar transverse charge transition densities can be defined for $J^P = \frac{3}{2}^+$ states such as the  $\Delta(1232)\frac{3}{2}^+$. This has been studied in~\cite{Carlson:2007xd} and is shown in Fig.~\ref{p-Delta-TCD}.

A comparison of $N(1440)\frac{1}{2}^+$ and $N(1535)\frac{1}{2}^-$ is shown in Figure~\ref{charge_densities}. There are clear differences in the charge transition densities between the two states. The Roper state has a softer positive core and a wider negative outer cloud than $N(1535)\frac{1}{2}^-$ and develops a larger shift in $b_y$ when the proton is polarized along the $b_x$ axis.  

\noindent
\subsubsection{Single Quark Transition Model}

Many of the exited states for which there is information about the transition form factors available have been assigned as members of the  $[SU(6), L^P]$ = $[70,1^-]$ super multiplet of the $[SU(6)\otimes O(3)]$ symmetry group. In a model, where only single quark transitions to the excited states are considered~\cite{Hey:1974qe,Cottingham:1978za,Burkert:2002zz}, only 3 of the amplitudes need to be known to determine the remaining 16 transverse helicity amplitudes for all states in $[70,1^-]$ including on neutrons. However, the picture is now more complicated due to the strong admixture of meson-baryon components to the single quark transition especially in the lower $Q^2$ range. This requires a model to separate the single quark contributions from the hadronic part before projections for other states can be made~\cite{Ramalho:2014pra}.

\subsection{Higher mass baryons and hybrid baryons} 
\begin{figure}[ht!]
\centerline{\resizebox{0.9\columnwidth}{!}{\includegraphics{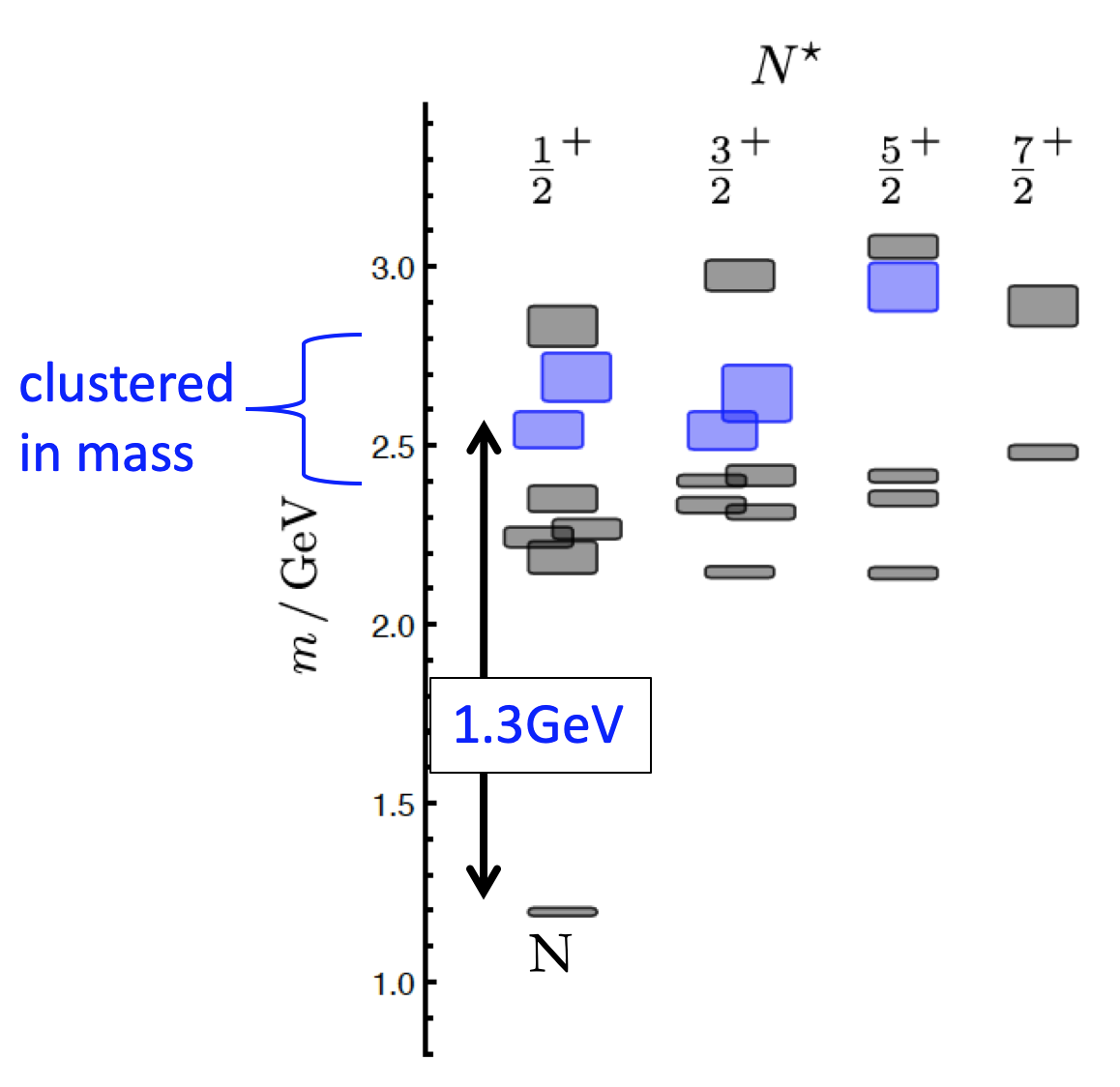}}}
\caption{Projections of excited baryons with dominant gluonic components (marked in blue shades) in LQCD with 400MeV pions. The lowest hybrid baryon is projected with mass 1.3 GeV above the nucleon mass. The $1/2^+$ and $3/2^+$ states are clustered in a narrow mass range of about 200 MeV.  }
\label{Hybrid-baryons}
\end{figure}

 The existence of baryons containing significant active gluonic components in the wave function has been predicted some decade ago~\cite{Dudek:2012ag} employing Lattice QCD simulations. The lowest such "hybrid" state is expected to be a $J^P=\frac{1}{2}^+$ nucleon state. LQCD projects a mass of 1.3 GeV above the nucleon mass, i.e. approximately 2.2-2.3~GeV, and several other states should appear close by in $J^P$ = $\frac{1}{2}^+$ and $J^P$ = $\frac{3}{2}^+$, as seen in Fig.~\ref{Hybrid-baryons}. 

How do we identify these states? Hybrid baryons have same spin-parity as other ordinary baryons. In contrast to hybrid mesons, there are no hybrid baryons with "exotic" quantum numbers. One possibility is to search for more states than the quark model predicts in some mass range. The other possibility is to study the transition form factors of excited states. Hybrid states may be identified as states with a different $Q^2$ behavior than what is expected from a 3-quark state. The sensitivity~\cite{Li:1991yba} is demonstrated for the Roper resonance that projected a very rapid drop of the $A_{1/2}(Q^2)$ with $Q^2$, and $S_{1/2}(Q^2) \sim 0$ prediction. Both are incompatible with what we know today about the Roper resonance.   
Precision electroproduction data in the mass range above 2 GeV will be needed to test high mass states for their potential hybrid character, e.g. from experiments at CLAS12~\cite{Lanza:2021ayj}.

\subsection{Conclusions and Outlook}

In this contribution we have focused on more recent results of nucleon resonance transition amplitudes and their interpretation within LQCD and within most advanced approaches, e.g. in light front relativistic quark models and approaches with traceable links to first principle QCD such as Dyson-Schwinger Equations \cite{Roberts:2007ji} and light cone sum rules~\cite{Braun:2009jy}.  
These calculations describe the transition form factors at $Q^2 \ge  2$~GeV$^2$, while at lower $Q^2$ values hadronic degrees of freedom must be included and could even dominate contributions of the quark core. 

For the lowest mass states, $\Delta(1232)\frac{3}{2}^+$ and the Roper $N(1440)\frac{1}{2}^+$, LQCD calculations have been carried out that are consistent with the data within large uncertainties. These calculations are about one decade old, and new data, with higher precision in more extended kinematic range have been added to the database that warrant new Lattice calculations at the physical pion mass to be carried out.

Over the past decade, eight baryon states in the mass range from 1.85 to 2.15 GeV have been either discovered or evidence for the existence of states has been significantly strengthened. Some of these states are in 
the mass range and have $J^{PC}$ quantum numbers that could have significant contributions of gluonic components. Such "hybrid" states are in fact predicted in LQCD~\cite{Dudek:2012ag}. These states appear with the same quantum numbers as ordinary quark excitations, and can only be isolated from ordinary states due to the $Q^2$ dependence of their helicity amplitudes~\cite{Li:1991yba}, which is expected to be quite different from ordinary 3-quark excitation. The study of hybrid baryon excitations then requires new electroproduction data especially at low $Q^2$~\cite{Lanza:2021ayj} with different final states and with masses above 2 GeV.  
Despite the very significant progress made in recent years to further establish the light-quark baryon spectrum and explore the internal structure of excited states and the relationship to QCD~\cite{Carman:2020qmb,Proceedings:2020fyd}, much remains to be done. A vast amount of precision data already collected needs  to be included in the multi-channel analysis frameworks, and polarization data are still to be analyzed. 
There are approved proposals to study resonance excitation at much higher $Q^2$ and with higher precision at Jefferson Lab with CLAS12~\cite{CLAS:2022kta,Burkert:2018nvj}, which may begin to reveal the transition to the bare quark core contributions at short distances.

A new avenue of experimental research has recently been opened up with the first data-based extraction of a gravitational property of the proton, its internal pressure distribution, which is represented by the gravitational form factor $D^q(t)$. It is one of the form factors of the QCD matrix element of the  energy-momentum tensor, its internal pressure and shear stress distribution in space~\cite{Burkert:2018bqq,Burkert:2021ith}. These properties, as well as the distribution of mass and angular momentum, and are accessible directly in gravitational interaction, which is highly impractical. However the relevant gravitational form factor $D^q(t)$ for the ground state nucleon can be 
accessed indirectly through the process of deeply virtual Compton scattering and in time-like Compton scattering~\cite{Ji:1996nm,CLAS:2021lky}. Both processes, having a $J=1$ photon in the initial state as well as in the final state, contain components of $J=2$ that couple to the proton through a tensor interaction, as gravity does~\cite{Polyakov:2018zvc}. 

Mechanical properties of resonance transitions have recently been explored for the $N(1535)\frac{1}{2}^- \to N(938)$ 
gravitational transition form factors calculations in~\cite{Ozdem:2019pkg} and in \cite{Polyakov:2020rzq}. In order to access these novel gravitational transition form factors experimentally, the nucleon to resonance transition generalized parton distributions must be studied. The framework for studying the $N \to N(1535)$ transition GPDs, which would enable experimental access to these mechanical properties, remains to be developed. The required effort is quite worthwhile as a new avenue of hadron physics has opened up that remains to be fully  explored.


\bibliography{main}
\end{document}